\documentclass[12pt, a4paper]{article}
\pdfoutput=1

\usepackage{jcappub}
\usepackage[utf8]{inputenc}

\newcommand{\taus}{s}
\newcommand{\mbf}[1]{\mathbf{#1}}
\newcommand{\SP}{{Schr\"odinger-Poisson }}

\def\be{\begin{equation}}
\def\ee{\end{equation}}

\def\bea{\begin{eqnarray}}
\def\eea{\end{eqnarray}}

\begin{document}

\begin{titlepage}
\thispagestyle{empty}

\rightline{DESY 19-176}
\rightline{TUM-HEP-1238/19}
\bigskip\

\vspace{1cm}
\begin{center}
{\fontsize{18}{24}\selectfont  \bfseries  {The \SP method for Large-Scale Structure\\ [12pt]  }} 
\end{center}
\begin{center}
{\fontsize{12}{18}\selectfont Mathias Garny$^{1}$, Thomas Konstandin$^{2}$ and Henrique Rubira$^{2}$} 
\end{center}

\begin{center}
\textsl{$^1$  \small Physik Department T31, James-Franck-Stra\ss e 1, \\ Technische Universit\"{a}t M\"{u}nchen, D–85748 Garching, Germany}
\vskip 8pt

\textsl{$^2$  \small Deutsches Elektronen-Synchrotron DESY,\\ Notkestra\ss e 85, D-22607 Hamburg, Germany}
\vskip 8pt

\end{center}

\vspace{1.2cm}
\hrule \vspace{0.3cm}
\noindent {\bf Abstract}\\[0.1cm]
We study the \SP (SP) method in the context of cosmological large-scale structure formation in an expanding background. 
In the limit $\hbar \to 0$, the SP technique can be viewed as an effective method to sample the phase space distribution of cold dark matter that remains valid on non-linear scales.
We present results for the 2D and 3D matter correlation function and power spectrum at length scales corresponding to the baryon acoustic oscillation (BAO) peak. 
We discuss systematic effects of the SP method applied to cold dark matter and explore how they depend on the simulation parameters. 
In particular, we identify a combination of simulation parameters that controls the scale-independent loss of power observed at low redshifts, 
and discuss the scale relevant to this effect.
\vskip10pt
\hrule
\vskip10pt

\end{titlepage}

\thispagestyle{empty}
\tableofcontents

\clearpage
\thispagestyle{plain}
\pagenumbering{arabic}
\setcounter{page}{1}

\section{Introduction}
\label{sec:introduction}

 The Large-Scale Structure (LSS) of the universe is a powerful cosmological probe, with current data from galaxy surveys
becoming competitive compared to Cosmic Microwave Background (CMB) measurements \cite{Aghanim:2018eyx} for certain parameters within the standard $\Lambda$CDM model, 
and also providing complementary information compared to the CMB in extended cosmological models 
\cite{Alam:2016hwk,Abolfathi:2017vfu,Abbott:2018jhe,Abbott:2018xao,Troster:2019ean}.
During the next decade an even higher level of precision around the Baryonic Acoustic Oscillation (BAO) 
peak will be reached with ground- and space-based surveys including Euclid \cite{Amendola:2012ys}, DESI \cite{Levi:2019ggs}, PFS \cite{Ellis:2012rn} and LSST \cite{Abell:2009aa}.

In view of the exquisite observational precision of future surveys, at or below the percent level for many LSS observables, it is timely to scrutinize
existing frameworks that are used to obtain theoretical predictions and also explore alternative approaches. Focusing on dark matter clustering, the
standard technique based on $N$-body simulations~\cite{Frenk:2012ph, Springel:2005mi} is being pushed to higher volumes and resolution, and refined
in order to increase the level of precision. Nevertheless, reaching percent accuracy with acceptable computational effort is not a trivial task \cite{Schneider:2015yka}.
The major alternative are well-known perturbation theory techniques and variants thereof~\cite{Scoccimarro:1995if,Bernardeau:2001qr,Crocce:2005xy,Carlson:2009it,
Baumann:2010tm,Rampf:2012up,Carrasco:2012cv,Blas:2013aba,Baldauf:2015zga,Konstandin:2019bay},
as well as effective descriptions such as
the halo model \cite{Cooray:2002dia}. 

The various approaches may be regarded as different  methods of sampling the phase space distribution of dark matter,
and obtaining (approximate) solutions of the Vlasov-Poisson equations that describe its evolution.
Inevitably, each method has its advantages and disadvantages. The collisionless fluid approximation breaks down after shell-crossing, 
but works well in the weakly non-linear regime. $N$-body simulations capture non-linear scales and sample regions of high
density very well. This is advantageous for studying, for example, dark matter halo properties and statistics. Nevertheless,
low-density regions are poorly sampled, while being interesting for probing e.g.~modifications of 
gravity \cite{Hamaus:2016wka,Voivodic:2016kog} and neutrino masses \cite{Massara:2015msa}. 
Furthermore, reconstructing higher moments of the distribution function, including
the velocity divergence and vorticity as well as the velocity dispersion tensor, requires to go beyond the 
standard $N$-body method~\cite{Hahn:2014lca,Buehlmann:2018qmm,Stucker:2019txm}.  
As a matter of principle, it is therefore desirable to investigate alternatives to the $N$-body technique.

One such alternative is provided by the \SP framework.
This approach is commonly considered in the context of ``fuzzy'' dark matter (FDM) 
-- an axion-like bosonic particle with mass $m\sim 10^{-22}\textrm{eV}$ \cite{Hu:2000ke} -- 
that has a macroscopic de Broglie wavelength $\hbar/mv$ of the order of kpc scales, leading
to a variety of distinct observational signatures, see e.g.~\cite{Hui:2016ltb}.
On cosmological scales, FDM is described by a condensate, that, in the non-relativistic limit and when neglecting any interactions apart from gravity,
obeys a Schr\"odinger equation governed by the gravitational potential that is self-consistently determined from the wave-function.

As proposed already by Widrow and Kaiser \cite{Widrow:1993qq}, the \SP (SP) system can also be viewed as a technique to sample 
the phase space distribution of cold dark matter, with a coarse-graining in phase space determined by $\hbar/m$. 
In this case, $\hbar/m$ plays a role similar to the number of particles $N$ in a $N$-body simulation, and
in the limit $\hbar/m\to 0$ one expects to recover a solution of the full Vlasov-Poisson system. This expectation
has been scrutinized and confirmed analytically and numerically
in various setups \cite{Uhlemann:2014npa,Schive:2014dra,Garny:2017xkc,Kopp:2017hbb,Mocz:2018ium,Veltmaat:2018dfz,Li:2018kyk,Uhlemann:2018gzz}. 
Here, a key point is that the SP system for the wave-function is solved directly, which, in contrast to the fluid-like Madelung representation, is free of
singularities for any value of $\hbar/m$.
A conceptual advantage of the SP method is that the coarse-grained six-dimensional phase space can be sampled with only two real-valued, three-dimensional 
functions (i.e. the real and imaginary part of the wave-function). Nevertheless, the SP method remains valid after shell-crossing and can capture the complex
features of the distribution function on non-linear scales. It is therefore particularly interesting for investigating higher moments of the
distribution function~\cite{Garny:2017xkc,Kopp:2017hbb}, and addressing questions like vorticity generation~\cite{Kopp:2017hbb,Mocz:2018ium,Li:2018kyk,Uhlemann:2018gzz}.

In this work, we discuss several systematic effects that control the accuracy
of the SP method when applied to cold dark matter clustering on cosmological scales in an expanding background. 
We investigate how they depend on the box size $L$ of the simulation volume, 
the number of points $N$ in each spatial dimension, the time step, as well as the coarse-graining parameter $\hbar/m$, and scrutinize the problem
of amplitude loss \cite{Li:2018kyk}.
The structure of the article is as follows: In section~\ref{sec:main_theory}, we set up our notations for the SP system and briefly review how the fluid, $N$-body and SP methods sample
the phase space distribution. In section~\ref{sec:numerical_implementation} we provide details about the numerical implementation.  
Section~\ref{sec:systematics} explores systematic effects in two dimensions, and in section~\ref{sec:3d_sp} we comment on the three-dimensional case
before concluding in section~\ref{sec:conclusion}. 
The appendices contain comments on the convergence of the SP code, the  energy conservation test, the initialization redshift, the computational time and the convergence in the one-dimensional case. 
  
\section{Sampling Vlasov phase space} \label{sec:main_theory}

The phase space distribution $f(\mbf{x},\mbf{p},\tau)$ for collisionless cold dark matter obeys the Vlasov equation~\cite{Bernardeau:2001qr} 
\begin{equation}\label{eq:Vlasov}
\frac{d f}{d\tau} = \frac{\partial f}{\partial \tau} + \frac{p_i}{am}\frac{\partial f}{\partial x_i} - am\nabla_i V \cdot \frac{\partial f}{\partial p_i} = 0 \, ,
\end{equation}
where $\tau$ is the conformal time, $\mbf{x}$ are comoving coordinates, and the gravitational potential $V$ for modes deep inside the horizon is given by the Poisson equation 
\begin{equation}\label{eq:poisson}
\Delta V = 4\pi G a^2 \bar{\rho}(\tau)\,\delta(\mbf{x},\tau) \, , \quad \int\frac{d^Dp}{(2\pi)^D} \, f(\mbf{x},\mbf{p},\tau) = 1+\delta(\mbf{x},\tau) \, .
\end{equation}
Here $D$ denotes the number of spatial dimensions, $\Delta=\nabla^2$ is the $D$-dimensional Laplace operator with respect to comoving coordinates, $\bar\rho(\tau)$ the average
matter density, and $\delta(\mbf{x},\tau)=\rho(\mbf{x},\tau)/\bar\rho(\tau)-1$ the density contrast.
The Vlasov equation~(\ref{eq:Vlasov}) is a non-linear partial differential equation in $2\times D+1$ dimensions, and hence quite hard to solve directly. 

In the following sections we briefly review how the phase space distribution $f$ is described in the
fluid approximation (section~\ref{sec:EP}), in $N$-body simulations (section~\ref{sec:nbody}), and via the \SP system (section~\ref{sec:SPmethod}). 
We stress again that in the limit $\hbar\to 0$, the SP method should be regarded as an alternative method to sample the (coarse-grained) phase space distribution
of cold dark matter, rather than a dual to the fluid system \cite{Widrow:1993qq,Uhlemann:2014npa,Garny:2017xkc,Kopp:2017hbb,Mocz:2018ium}. 
In particular, the map between the fluid and the SP description through the Madelung representation contains singularities in the limit $\hbar \rightarrow 0 $ and fails after shell-crossing. While the fluid approximation breaks down, the SP system is free of singularities also after shell-crossing.

\subsection{Euler-Poisson (EP)} \label{sec:EP}

The Vlasov equation can be converted into a coupled set of equations for the cumulants of the phase space distribution
in momentum space. The generating function for the cumulants is given by
\be
  \exp\left[{\cal C}(\mbf{x},\mbf{l},\tau)\right]\equiv \int \frac{d^Dp}{(2\pi)^D}\,\exp\left[i\mbf{l}\cdot\frac{\mbf{p}}{am}\right]\,f(\mbf{x},\mbf{p},\tau)\,.
\ee
The lowest order cumulants are related to the density contrast $\delta$, the bulk velocity field $\mbf{u}$ and the velocity dispersion $\sigma_{ij}$ by
\be\label{eq:Cl}
  {\cal C}|_{\mbf{l}=0}=\ln(1+\delta), \quad \nabla_\mbf{l}{\cal C}|_{\mbf{l}=0}=\mbf{u}, \quad \nabla_{\mbf{l}_i}\mbf{\nabla}_{\mbf{l}_j}{\cal C}|_{\mbf{l}=0}=\sigma_{ij}\,.
\ee
The Vlasov equation \eqref{eq:Vlasov} yields the equation of motion for the generating function~\cite{Pueblas:2008uv}
\be
  \frac{\partial{\cal C}}{\partial\tau}+aH(\mbf{l}\cdot\nabla_\mbf{l}){\cal C}+\nabla{\cal C}\cdot\nabla_\mbf{l}{\cal C}+(\nabla\cdot\nabla_\mbf{l}){\cal C}=-\mbf{l}\cdot\nabla V\,.
\ee
By Taylor expanding in $\mbf{l}$ one obtains a coupled hierarchy of equations for the cumulants, with the lowest two being the familiar continuity and Euler equations~\cite{Bernardeau:2001qr}
\bea
    \partial_\tau \delta &=& - \nabla\cdot [(1+\delta )\mbf{u}] \, ,\\
   \partial_\tau \mbf{u}_i + aH\mbf{u}_i + (\mbf{u}\cdot \nabla) \mbf{u}_i & =& - \nabla_i V -\frac{1}{1+\delta}\nabla_j[(1+\delta)\sigma_{ij}]\,.
\eea 
The Euler equation depends on $\sigma_{ij}$. Its equation of motion is obtained from the second-order Taylor expansion of \eqref{eq:Cl}, and in turn depends on the
third cumulant due to the last two terms on the left-hand side of \eqref{eq:Cl}. Proceeding further, one obtains a coupled hierarchy of equations for the cumulants. 

The perfect pressureless fluid (PPF) approximation corresponds to neglecting $\sigma_{ij}$. The perturbative solution of the continuity and Euler equations leads to
the well-known Standard Perturbation Theory (SPT). 
In terms of the generating function, the perfect pressureless fluid approximation corresponds to the ansatz ${\cal C}=A+\mbf{l}\cdot\mbf{B}$, with a linear dependence on $\mbf{l}$.
It can be readily checked that this ansatz indeed provides a self-consistent solution of \eqref{eq:Cl}, and therefore of the full Vlasov equation.
This particular class of solutions corresponds to the phase space distribution given by a $D$-dimensional
hypersurface in $2D$ phase space,
\begin{equation}
  f_{PPF}(\mbf{x},\mbf{p},t) = (1+\delta(\mbf{x},\tau))(2\pi)^D\delta^{(D)}[\mbf{p}-a m\mbf{u}(\mbf{x},\tau)]\,,
\end{equation}
that describes a single stream of dark matter particles. Therefore, the PPF approximation has to break down once shell-crossing occurs.
Formally, the density contrast would become singular at the space-time location of the first shell-crossing within the PPF approximation,
such that the PPF solution of the full Vlasov equation cannot be continued to later times. Instead, non-zero velocity dispersion $\sigma_{ij}$
as well as higher order cumulants are generated
in regions with multiple streams. For realistic initial conditions, shell-crossing occurs first on the smallest scales, while larger scales are still close to
the single-stream regime. This motivates the Effective Field Theory (EFT) approach that consists of a fluid description for large-scale modes, complemented with
an effective expression for $\sigma_{ij}$ on the right-hand side of the Euler equation. At lowest order, the effective velocity dispersion tensor can be parameterized by 
an effective pressure and viscosity as well as possibly a stochastic noise component 
\cite{Baumann:2010tm,Mercolli:2013bsa,Abolhasani:2015mra,Floerchinger:2016hja}. 
For an approach using a truncation of the hierarchy at the third order, see \cite{Erschfeld:2018zqg}.

\subsection{$N$-body} \label{sec:nbody}

The $N$-body simulation technique can be regarded as a sampling of the $2\times D$-dimensional dark matter phase space distribution with discrete point particles with mass $m$. 
For a simulation with comoving box size $L$ and with $N_{\rm bodies}$ particles, the mass of the hypothetical point particles is chosen such that 
$m N_{\rm bodies}/L^D=\bar\rho(t)a^D$.
The corresponding Klimontovich phase space distribution has the form 
\begin{equation}
f_K(\mbf{x},\mbf{p},\tau) = \frac{m}{\bar{\rho}} \sum_{i=0}^{N_{\rm bodies}} \delta^{(D)}[\mbf{x}-\mbf{x}_i(\tau)]\, (2\pi)^D\delta^{(D)}[\mbf{p}-\mbf{p}_i(\tau)]\, .
\end{equation}
It is expected to approximate the continuous phase space distribution $f$ in the limit $N_{\rm bodies}\rightarrow\infty$. 
By construction, the resolution of $f_K$ is higher in the densest regions but very poor in void regions. 
This sampling is advantageous when studying the distribution and properties of dark matter halos,
but makes it more challenging to reconstruct, for example, the velocity field or the velocity dispersion.
In addition, warm dark matter models with a suppressed linear power spectrum are challenging due
to artificial structures on small scales related to the discreteness of the phase space sampling~\cite{Schneider:2014rda}. For an extension of the
$N$-body technique based on  phase space interpolation, see refs.~\cite{Hahn:2014lca, Buehlmann:2018qmm, Stucker:2019txm}.

\subsection{\SP (SP)}
\label{sec:SPmethod}

The Schr\"odinger equation 
\begin{equation}\label{eq:schrodinger}
    i\hbar \partial_\tau \psi = -\frac{\hbar^2}{2am}\Delta\psi + amV\psi\,,
\end{equation}
with the potential given by 
\begin{equation} 
    \Delta V = 4\pi G\bar\rho(\tau)a^2(|\psi|^2 -1)\,,
\end{equation}
describes a classical bosonic condensate interacting through gravity. One can also think of this condensate as a superfluid, mapped by the Madelung representation 
\begin{equation}\label{eq:Madelung}
\psi = \sqrt{1+\delta}\, e^{i\phi/\hbar} \, ,
\end{equation} 
such that the density (normalized to the average density) is given by  $\rho/\bar\rho \equiv |\psi|^2$.
The phase $\phi$ represents a velocity potential, with bulk velocity given by $\mbf{u} \equiv \nabla \phi/(am)$.  From this definition, one finds
\begin{equation} \label{eq:spvelocity}
\mbf{u} = \frac{i\hbar}{2(1+\delta) \, am}[(\nabla\psi^*)\psi - (\nabla\psi)\psi^*] \, ,
\end{equation}
and we can recover the fluid equations:
\bea
    \partial_\tau \delta &=& - \nabla\cdot [(1+\delta )\mbf{u}] \, ,\\
   \partial_\tau \mbf{u} + aH\mbf{u} + (\mbf{u}\cdot \nabla) \mbf{u} & =& - \nabla V + \frac{\hbar^2}{2a^2m^2} \nabla \left(\frac{\Delta \sqrt{\rho}}{\sqrt{\rho}}\right)\,.
\eea 
The last term is often called ``quantum pressure'' and is not present in the perfect pressureless fluid equations described in section~\ref{sec:EP}. This term -- at least at linear order\footnote{At higher orders in wave-perturbation theory, the quantum pressure can act with the same sign as gravity \cite{Li:2018kyk}.} -- prevents structure formation at very small scales. 
Comparing to the gravitational force term, this corresponds to a (comoving) Jeans scale~\cite{Khlopov:1985jw, Hu:2000ke} (see also \cite{Hui:2016ltb})
\be\label{eq:Jeans}
  k_J = \frac{2(\pi G\bar\rho a^2)^{1/4}(am)^{1/2}}{\hbar^{1/2}} = (6\Omega_m)^{1/4}\sqrt{\frac{a^2Hm}{\hbar}} \,.
\ee
This can be understood as a consequence of the Heisenberg uncertainty principle, in which at physical (not comoving) scales smaller than $\sqrt{\hbar/(mH)}$ the velocity dispersion increases. This effect in our result will become evident in section~\ref{sec:falling}.

Using the Madelung transformation~(\ref{eq:Madelung}), one may be tempted to claim that the usual PPF fluid equations are recovered in the limit $\hbar \rightarrow 0$. This affirmation is somewhat simplistic, since the Madelung transformation  leads to singularities in the limit $\hbar \rightarrow 0$ and an ambiguity as $\rho \rightarrow 0$. Typically, the wave-function develops strongly
oscillatory features shortly after shell-crossing, including space-time points where both the real and imaginary parts vanish such that $\rho \rightarrow 0$. 
This implies that the mapping of the wave-function to a fluid description becomes ambiguous after shell-crossing.
However, this does not affect the description based on the wave-function $\psi$, which remains valid throughout the shell-crossing regime.

We adopt here the point-of-view that the SP system is not a dual of the fluid model but an alternative method for sampling the (coarse-grained) phase space of dark matter 
\cite{Widrow:1993qq,Uhlemann:2014npa,Garny:2017xkc,Kopp:2017hbb,Mocz:2018ium}. 
The properties of the SP method for this purpose are different from the characteristics of the EP or $N$-body approach, as we discuss below, and may, therefore, allow to address
different questions compared to conventional techniques.
When employing the SP method to describe cold dark matter, the value of $\hbar$ should be regarded as an effective parameter that controls the resolution in phase space.
In this sense, $\hbar$ is on the same footing as the parameters that are related to the discrete sampling in 
$N$-body simulations (i.e. the $N$-body particle mass $m$ and force softening length; these parameters are not required for the SP method).

To simplify the notation hereafter, we define a rescaled potential and a parameter $\kappa$ that is related to the Jeans scale \eqref{eq:Jeans}  \cite{Garny:2017xkc},
\begin{equation} \label{eq:short_definitions}
\kappa(t) = \frac{\hbar}{a^2mH(t)} \quad \textrm{and} \quad \bar{V} = \frac{mV}{\hbar H}\,.
\end{equation}
Furthermore, for simplicity, we assume a background cosmology described by a matter-dominated universe, with $\bar\rho=\rho_0/a^3$ and Friedmann equation
\begin{equation}
H^2 = \frac{8\pi G}{3a^3}\rho_0\,.
\end{equation}
Changing the time variable to $\eta = \ln a$, we can write the Schr\"odinger and the  Poisson equation as
\bea
\label{eq:sp_new}
    i\partial_\eta \psi &=& -\frac{\kappa(\eta)}{2}\Delta\psi + \bar{V}\psi \, ,  \\
\label{eq:vbar}
    \Delta \bar{V} &=& \frac{3}{2\kappa(\eta)}(|\psi|^2 -1) \,.
\eea
The cosmic background expansion enters only via the time-dependence of $\kappa$. Note that a static background could be described by the same set of equations, with constant $\kappa$.
In the case of a matter-dominated universe one finds
\begin{equation}
\kappa(\eta) = \frac{\hbar}{m} \frac{1}{a^2H(\eta)} = \frac{\hbar}{mH_0} \frac{1}{a^{1/2}}   = \frac{\hbar}{mH_0} \exp{\left(-\eta/2\right)}\,,
\end{equation}
where $\kappa_0$ is the value of $\kappa$ today. Notice that $\hbar$ enters explicitly only via $\kappa$. 

\subsubsection*{Reconstructing the phase space in SP}

In order to reconstruct the phase space density  distribution from the wave function, one can perform a Wigner transformation
\begin{equation}
f_W(\mbf{x}, \mbf{p},\tau) = \frac{1}{(\pi\hbar)^3}\int d^3x' \exp{\left[ 2\frac{i}{\hbar}\mbf{p}\cdot\mbf{x'}\right]} \psi(\mbf{x}-\mbf{x}',\tau)\psi^*(\mbf{x}+\mbf{x}',\tau)\, .
\end{equation}
Since the Wigner phase space distribution can assume negative values and features oscillations on $\hbar$ scales, its relation with the classical (Vlasov) distribution function is deficient \cite{Uhlemann:2014npa}. One can instead filter both classical and Wigner distribution functions on these scales, making the correspondence of the coarse-grained versions in phase space evident \cite{Takahashi:distributionfunctions}. This is equivalent to eliminating quantum uncertainties (see e.g.~figure 1 of \cite{Takahashi:distributionfunctions}). 
By convoluting the Wigner distribution with a Gaussian kernel in both position and momentum space, with widths $\sigma_x$ and $\sigma_p$, respectively,
one obtains a non-negative result if $\sigma_x\sigma_p\geq \hbar/2$. If this inequality is saturated, the coarse-grained distribution function can be
constructed in a simpler way via the
Husimi transformation
\begin{equation}
\psi_H(\mbf{x},\mbf{p},\tau;\sigma_x) = \int d^Dy K_H(\mbf{x},\mbf{y},\mbf{p};\sigma_ x) \psi(\mbf{y},\tau) \, ,
\end{equation}
with the kernel 
\begin{equation}
\quad K_H(\mbf{x},\mbf{y},\mbf{p};\sigma_x) = \frac{\exp{\left[ \frac{-|\mbf{x}-\mbf{y}|^2}{4\sigma_x^2} - \frac{i\mbf{p}\cdot \mbf{y}}{\hbar} \right]}}{(2\pi\hbar)^{\frac{D}{2}}(2\pi\sigma_x^2)^{\frac{D}{4}}} \, ,
\end{equation}
such that the final distribution is given by
\begin{equation}
f_H(\mbf{x},\mbf{p},\tau) = |\psi_H|^2 \, ,
\end{equation}
which is non-negative by construction. 

\section{Numerical implementation} \label{sec:numerical_implementation}

In this section, we discuss the algorithm used to solve the SP equations and for setting up the initial conditions. 
After studying the impact of the initial redshift, we discuss the evolution of the density field and the power spectrum in 2D.

\subsection{Initial conditions}\label{sec:ic}

For setting up the initial conditions for the wave function $\psi$ we use the Madelung representation \eqref{eq:Madelung}, which is valid up to shell-crossing and should be broken only at low redshifts
for the scales resolved in our simulation.
The complex phase $\phi$ is set by the velocity potential (\ref{eq:spvelocity}). To set its initial condition, we use the Zel'dovich approximation \cite{Zeldovich:1969sb} 
\begin{equation}
\frac{\phi}{\hbar} = -\frac{\delta}{\kappa\Delta } \, .
\end{equation}
To reduce cosmic variance, we initialize the density contrast $\delta$ with the absolute value given by the square root of the power spectrum \cite{Angulo:2016hjd}, and a random phase $\zeta$,
\begin{equation}
\label{eq:Ps_initial}
\delta(\mbf{k}) = \sqrt{P(k) \, L^D}e^{i\zeta(\mbf{k})} \, .
\end{equation}

For the initial conditions in the 1D and 2D cases we define 
\begin{equation}
P_{1D} = \frac{k^2}{2\pi}P_{3D} \quad \textrm{and} \quad P_{2D} = \frac{k}{\pi}P_{3D} \, ,
\end{equation}
where $P_{3D}$ denotes the linear power spectrum for a $\Lambda$CDM cosmology generated by Boltzmann solvers like CAMB \cite{Lewis:2002ah} or CLASS \cite{Lesgourgues:2011re}. 
The definition above ensures that the ($n$-th) direction-independent moments of the power spectrum are the same in the different dimensions
\be
\int_{-\infty}^\infty \frac{dk}{(2\pi)} \, k^n\, P_{1D} \, 
 = \int \frac{d^2k}{(2\pi)^2} \, k^n\, P_{2D} \, 
 = \int \frac{d^3k}{(2\pi)^3} \, k^n\, P_{3D} \, .
\ee
With this normalization, the BAO peak in the linear correlation function has comparable numerical values in the case of 1D, 2D and 3D. For the linear input spectrum, we used a $\Lambda$CDM cosmology with parameters as in \cite{Kim:2011ab}.
The $D$-dimensional matter correlation function is then given by
\begin{equation}
\xi(x) = \int \frac{d^Dk}{(2\pi)^D} \phantom{1}  e^{i\mbf{x}\cdot \mbf{k}}P  \,.
\end{equation}
Explicitly, for 1D, 2D and 3D in the isotropic case one finds
\bea
\xi_{1D}(x) &=& \int \frac{dk}{(2\pi)}\phantom{1}  e^{ixk}P_{1D}  \,,\\
\xi_{2D}(x) 
&=& 2\pi \int_0^\infty dk \phantom{1}  J_0(xk)kP_{2D} = \frac{1}{(2\pi)^2} \mathcal{H}_0[P_{2D}]  \,,\\
\xi_{3D}(x) 
&=&   \frac{2}{(2\pi)^2} \int^{\infty}_{0} dk\phantom{1}  \frac{\sin{kx}}{kx} k^2P_{3D} = 
- \frac{1}{(2\pi)^2}\frac{1}{x}\operatorname{Im}\{\textrm{FT}(kP_{3D})\}  \,, 
\eea
where $J_0$ is the zeroth Bessel function and $\mathcal{H}_0$ is the Henkel transform of zeroth order.

\subsection{SP algorithm} \label{sec:sp_algo}

In this section, we explain the numerical algorithm used for solving the \SP equations. 
The time evolution in terms of the (time-ordered) Hamiltonian operator ${\cal H}=-\frac{\kappa}{2}\Delta + \bar{V}$ is given by
\bea 
\label{eq:rotation_psi}
\psi(x,\eta+\Delta \eta) &=& {\cal T} \, e^{ -\frac{i}{\hbar}\int_\eta^{\eta+\Delta\eta} \, {\cal H} \, d\eta}\psi(x,\eta) \nonumber \\
 &=& {\cal T} \, e^{ -i \int_\eta^{\eta+\Delta\eta} \left(-\frac{\kappa}{2}\Delta\psi + \bar{V}\psi \right) d\eta }\psi(x,\eta) \, ,
\eea
where ${\cal T}$ denotes time-ordering.
Before proceeding, it is convenient to remove the time dependence in the kinetic term by changing to a new time variable $\taus$ that is defined via $d\eta/d\taus = \kappa$. Next, for a small time step $\Delta \taus$, this can be written in term of the 
rotation operators (using leapfrog integration)
\be
\psi(x,\taus+\Delta \taus) = U_K(\Delta\taus/2) \, U_V(\Delta\taus) \, U_K(\Delta\taus/2)  \, \psi(x,\taus)\, ,
\ee
where we defined 
\be
\label{eq:rotations}
U_K(\Delta\taus) = \exp{\left(-\frac{i}{2}\Delta\taus \, k^2\right)} \, ,
\ee
and
\be
U_V(\Delta\taus) = \exp{\left( -i\int_\taus^{\taus+\Delta\taus} (\bar{V}/\kappa) \,  d\taus \right)} \, .
\ee
The leapfrog integration produces in principle an error of order $\left[ U_K,\left[ U_K,U_V \right]\right]$, in case 
the operators $U_K$ and $U_V$ are evaluated at the correct order. In particular, the integral in the potential term has to be calculated to the order $O(\Delta \taus^2)$. The time dependence of $|\psi^2|$ is at early times moderate such that the main time dependence in $U_V$ results from the explicit factor $1/\kappa^2$ (notice $\bar V \propto 1/\kappa$ in (\ref{eq:vbar})).

However, there is another constraint that one has to fulfill, coming from the time ordering of the Hamilton operator. In practice, we find that one has to limit the maximal angle $\theta_{\rm max}$ that can occur in a time step in arg$(U_K)$ or arg$(U_V)$. For $\theta_{\rm max} \lesssim 0.4$ the precision of the final $\psi$ indeed scales as $\Delta \taus^2$ (or $\theta_{\rm max}^2$) and the error in $\psi$ is beyond what we require (of order $10^{-5}$). For larger values of $\theta_{\rm max}$, however, the precision quickly deteriorates and the error in $\psi$ becomes of order unity. In summary, leapfrog integration improves precision greatly, but unfortunately this does not translate into a faster algorithm compared to a simple integration.   In this work we used $\theta_{\rm max} = 0.1$. For this choice, the main discretization error comes from spatial discretization. See appendix~\ref{app_convergence} for more details about the convergence of the wave-function.

In appendix~\ref{app_init}, we investigate the dependence on the initial redshift. For our fiducial choice of parameters, we find that the power spectrum changes by
less than $\sim 2$\% when varying $z_{\rm ini}$ between $50$ and $150$, which is compatible with the expected sensitivity for Zel'dovich initial conditions in $N$-body simulations.

\subsection{Density field evolution}

\begin{figure}[ht]
\centering
  \includegraphics[width=0.45\textwidth]{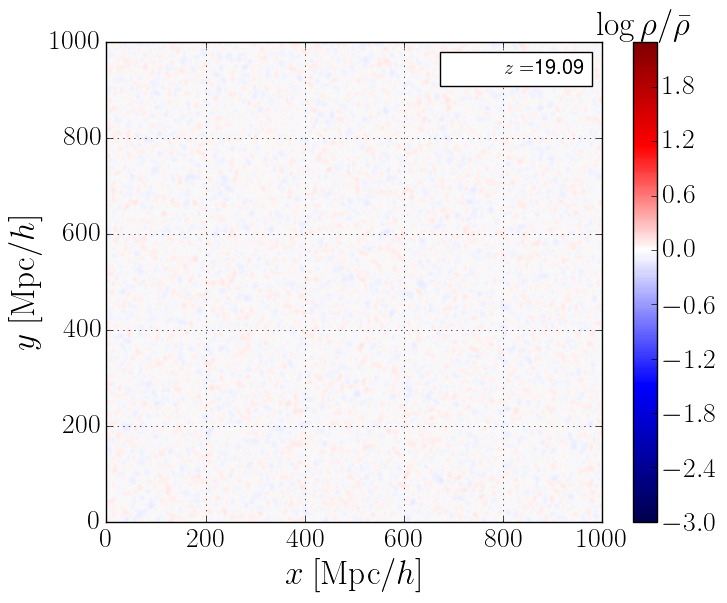}
  \includegraphics[width=0.45\textwidth]{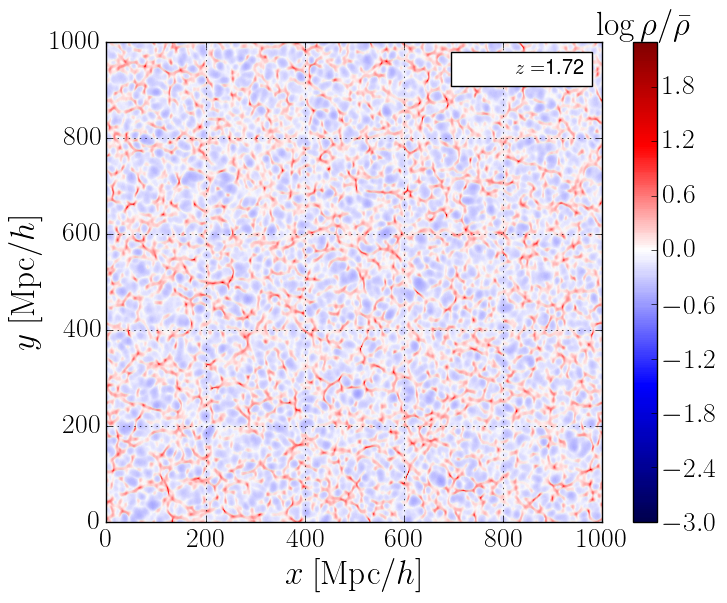}  
  \includegraphics[width=0.45\textwidth]{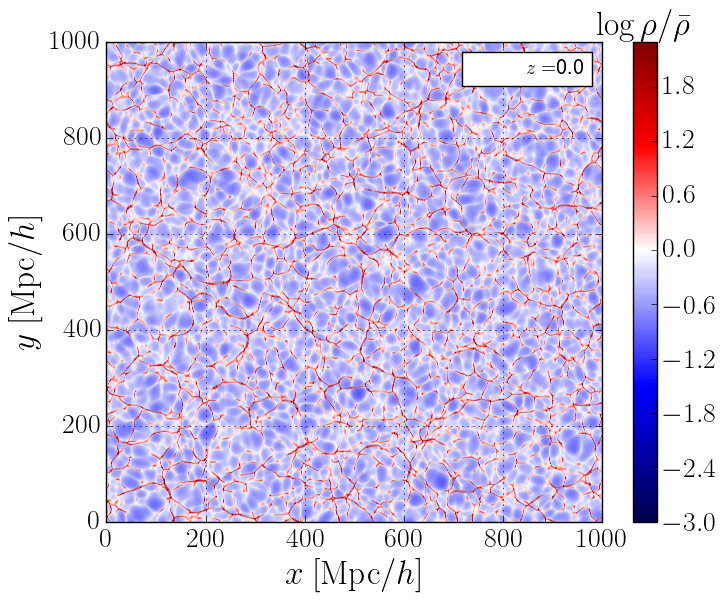}  
  \includegraphics[width=0.45\textwidth]{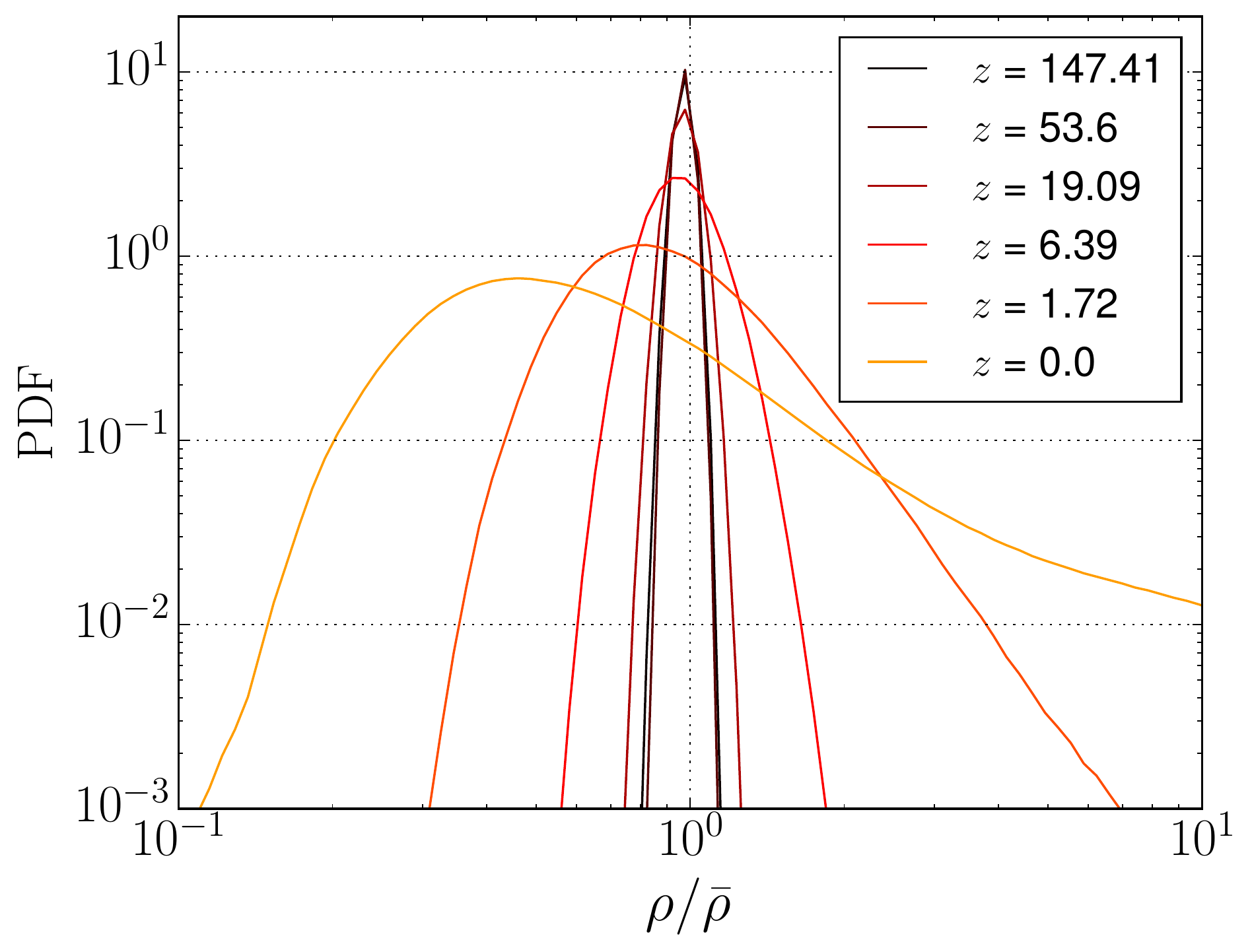}
\caption{\label{fig:dens_2d_z}%
\small Density contrast at three different redshifts ($z=19.09$, $z=1.72$ and $z=0$) for the 2D \SP system for $L=1000$~Mpc/h, $\kappa_0 = 1$ Mpc$^2/h^2$ and $N = 8192$. 
In the bottom-right, the density PDF at various redshifts is shown.}
\end{figure}

We evolve the 2D wave function in time starting from $z=147$ on a box with comoving side length $L=1000$~Mpc$/h$, $N = 8192$ grid points in each dimension, and $\kappa_0 = 1$ Mpc$^2/h^2$.
In figure~\ref{fig:dens_2d_z} we show the density contrast $\delta$ for the 2D SP system at three different redshifts. At $z\sim 19$, the density fluctuations are still almost Gaussian and $\delta$ is small. At $z=1.72$, structures become visible and, at $z=0$, one may recognize a ``cosmic web''. We also display the density PDF in figure~\ref{fig:dens_2d_z}, confirming the growth of non-linear structures. For the PDF we apply a top-hat filter in position space with smoothing length $2\,$Mpc$/h$. For a given set of simulation parameters, the PDF without filtering has
similar shape as the smoothed one, apart from the high-density tail.

\begin{figure}[ht]
\centering
  \includegraphics[width=0.45\textwidth]{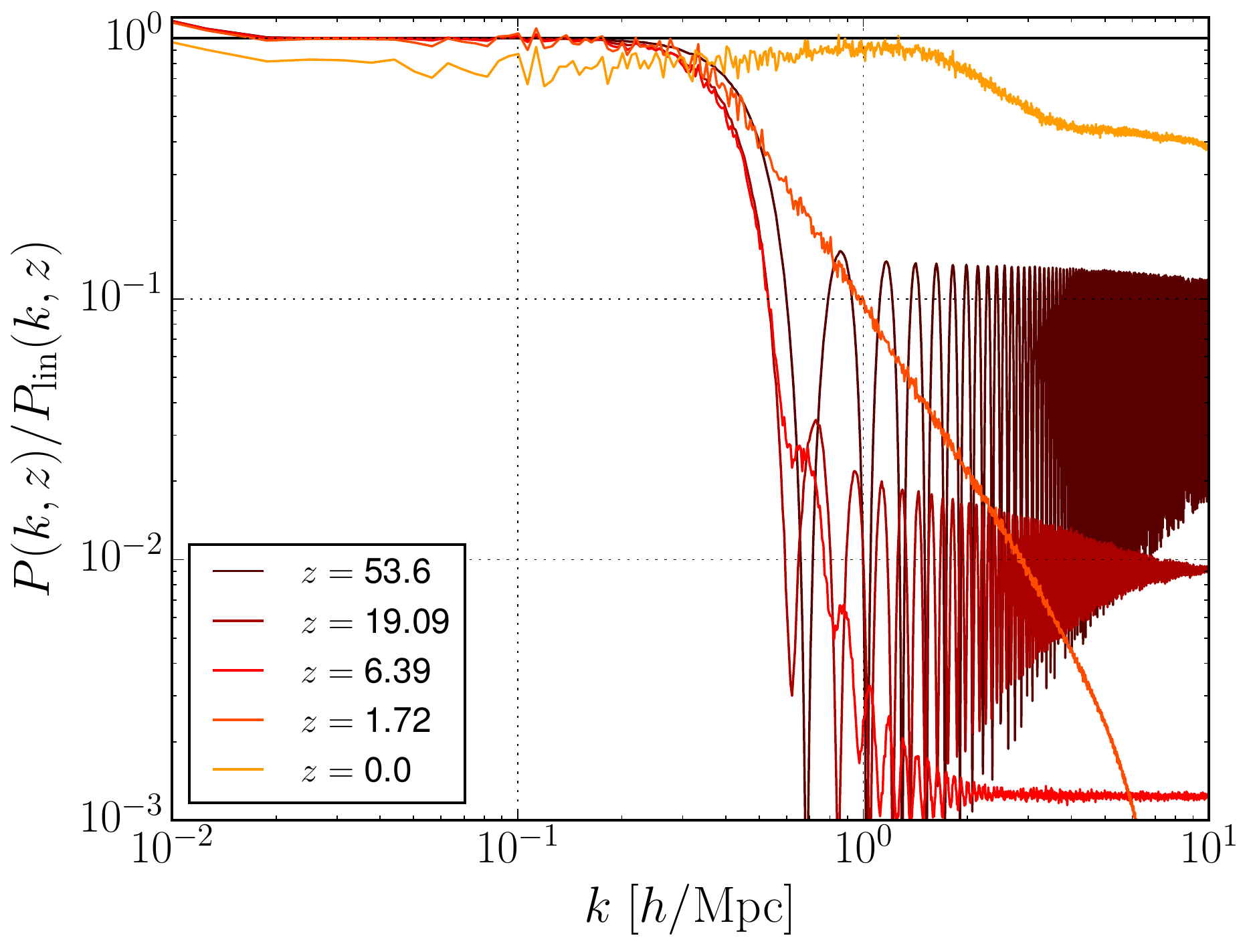}  
\caption{\label{fig:pdf_z_2d}%
\small Matter power spectrum divided by the linear power spectrum at redshifts $z=0,1.72,6.39,19.09,53.6$. 
}
\end{figure}

The matter power spectrum is shown in figure~\ref{fig:pdf_z_2d}. The time evolution of the SP system imprints three different types of features on the power spectrum, which we list below: 
\begin{enumerate}
\item A strong exponential (Jeans) suppression at small scales (section~\ref{sec:falling});
\item Sampling noise on large scales that were not present in the initial conditions (section~\ref{sec:noise});
\item A slight loss of power for all modes at low redshift (section~\ref{sec:amplitude}).
\end{enumerate}

The first effect is a physical property in fuzzy dark matter models related to the Jeans scale \eqref{eq:Jeans}, but should be considered as a systematic limitation
when applying the SP method to describe the phase space evolution of cold dark matter. 
Notice that, at late time, the Jeans scale does not suppress all power on small scales. So non-linear growth seems to 
be less affected than one would expect from the linear analysis of the system.
As shown below, the second item is essentially analogous to sampling noise in $N$-body simulations, which is
related to the finite number of modes. The third feature has already been recognized in the context of fuzzy dark matter~\cite{Li:2018kyk}, and is a systematic error of the (discretized) SP method
for both fuzzy and cold dark matter.


\section{The systematics of the \SP method}\label{sec:systematics}

In this section, we quantify each one of the three systematics effects mentioned above and evaluate their dependence on the simulation parameters, including the box size $L$, the number of lattice points in each dimension $N$, and the phase space resolution controlled by the value of $\hbar$. Since the latter enters in the rescaled SP equations (\ref{eq:sp_new}) only via the function $\kappa(\eta)$ (see \eqref{eq:short_definitions}),
we trade $\hbar$ for $\kappa_0$, the present value of $\kappa$.
We study variations around the fiducial values $L=1000$~Mpc$/h$, $N = 4096$ and $\kappa_0 = 1$ Mpc$^2/h^2$, which we found to be parameters that describe the BAO peak reasonably well while requiring a feasible amount of computational time (see appendix~\ref{app:time}). Furthermore, we use a fixed initial redshift $z=147$. 
For comparison, the original work \cite{Widrow:1993qq} used $N=256$ and $L=150$~Mpc$/h$ in 2D.

\begin{figure}[ht]
\centering
  \includegraphics[width=0.3\textwidth]{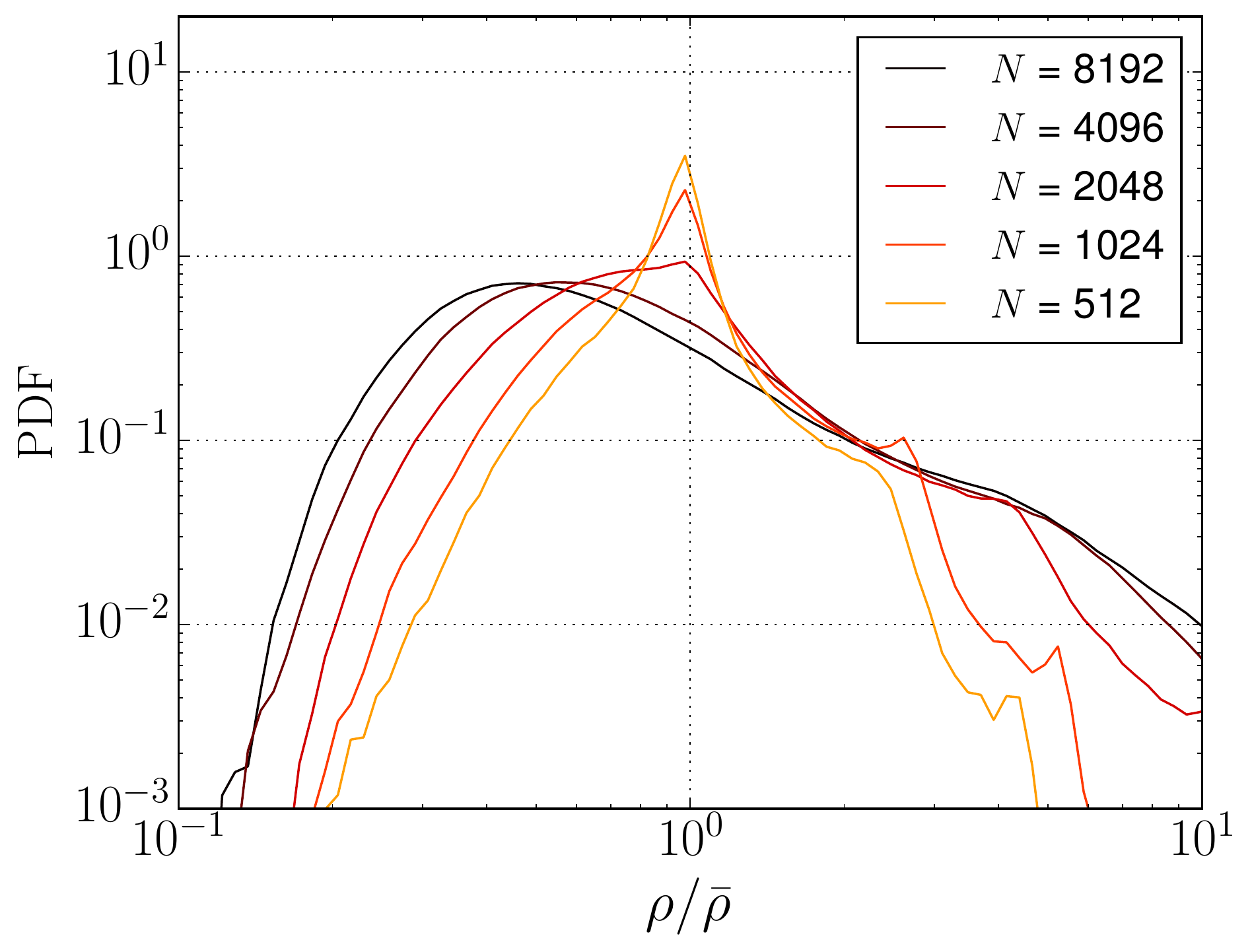}
  \includegraphics[width=0.3\textwidth]{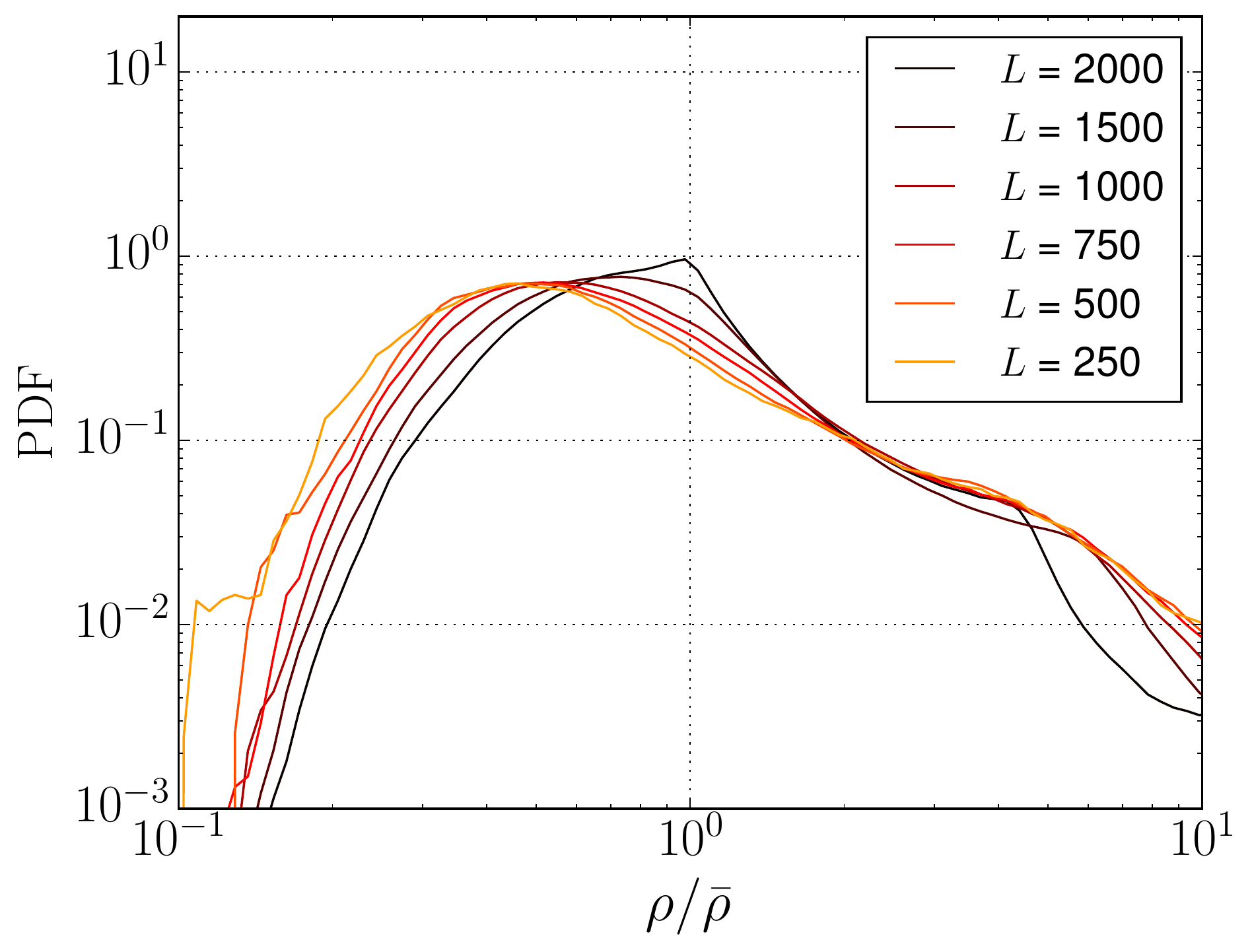}
  \includegraphics[width=0.3\textwidth]{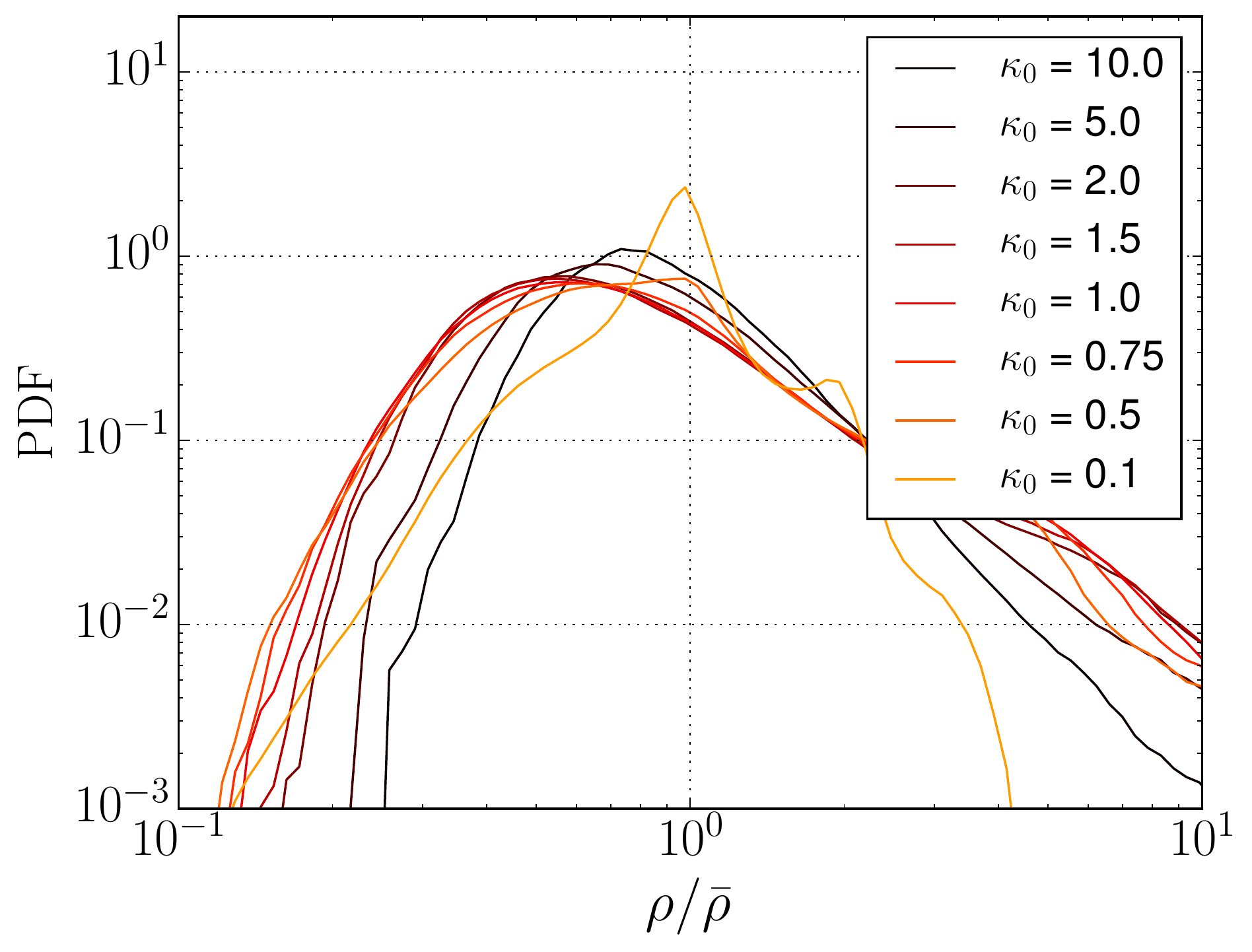}
\caption{\label{fig:pdfs_param_2d}%
\small Dependence of the PDF of the matter density on the simulation parameters:  For different grid sizes $N$ on the left; for different box sizes $L$ (in units of Mpc/$h$) in the middle, and for different $\kappa_0\propto\hbar $ (in units of Mpc$^{2}/h^2$) on the right. We use a top-hat filter in position space with a smoothing scale of $2\,$Mpc$/h$.}
\end{figure}

In figure~\ref{fig:pdfs_param_2d}, we show the dependence of the PDF on the simulation parameters ($N$, on the left; $L$ in the middle and $\kappa_0$ on the right). 
In the strict limit of (infinitely) \emph{cold} dark matter, the PDF is not expected to converge uniformly (without any coarse-graining) when increasing $N/L$, since smaller and smaller structures are resolved  \cite{Stucker:2017nmi}.
However, for the \SP system with fixed $\hbar$ (i.e. fixed $\kappa_0$), the Jeans scale acts as a coarse-graining length which should improve convergence. 
We find that increasing the number of lattice points enhances the deviation from a Gaussian distribution, and increases the tails of the PDF.
Nevertheless, for $N = 8192$ the PDF did not converge yet. As mentioned before, for the PDF we apply a top-hat filter in position space with smoothing length $2\,$Mpc$/h$ (see also appendix~\ref{app:1d}, in which we explore the convergence for the 1D case).
Decreasing the box size $L$ with $N$ fixed also improves the resolution of the non-linear modes, while increasing the box size too much leads to a loss of resolution. Decreasing $\kappa_0$ leads to similar effects that are accompanied by an overall loss of power in the fluctuations. Larger values of $\kappa_0$, in turn, increase the quantum pressure what also suppresses non-linearities. 
For intermediate values of $\kappa_0$ the result is relatively stable. 
While no clear picture emerges for the PDFs, the role of the different parameters will become more clear in the following when we study the power spectrum.

\subsection{Jeans suppression} \label{sec:falling}

The exponential loss of power at some scale $k_{\rm fall}$ related to the Jeans scale \eqref{eq:Jeans} is a characteristic property of the SP system. The Heisenberg uncertainty principle inhibits the formation of structures that are smaller than the Jeans scale. 
In the context of using the SP method to describe cold dark matter, the Jeans suppression has to be considered as a source of systematic errors.
In the left panel of figure~\ref{fig:pdf_z_2d}, we can see that shortly after initializing the simulation, the Jeans suppression strongly affects the power spectrum above around $\sim 1h/{\rm Mpc}$. To quantify the scale $k_{\rm fall}$ where the exponential suppression appears, we define it to be the largest mode for which the ratio of the power spectrum to the 
corresponding linearly evolved $\Lambda$CDM input power spectrum $P_{lin}(k,z)$ is smaller than $90\%$ 
\begin{equation} \label{eq:kfall}
k_{\rm fall} = \min(k) \quad \textrm{for which} \quad \frac{P(k,z_{ref})}{P_{lin}(k,z_{ref})} < 0.9  \,.
\end{equation}
We measure this scale at $\eta = -4$ ($z_{ref}=53.6$), when the other systematic effects are still irrelevant and the
system  already had enough time to develop Jeans suppression after being initialized with a $\Lambda$CDM spectrum at $z=147$. 
For the fiducial simulations used here we find
$k_{\rm fall} \simeq 0.3h/\textrm{Mpc}$.

\begin{figure}[ht]
\centering
  \includegraphics[width=0.45\textwidth]{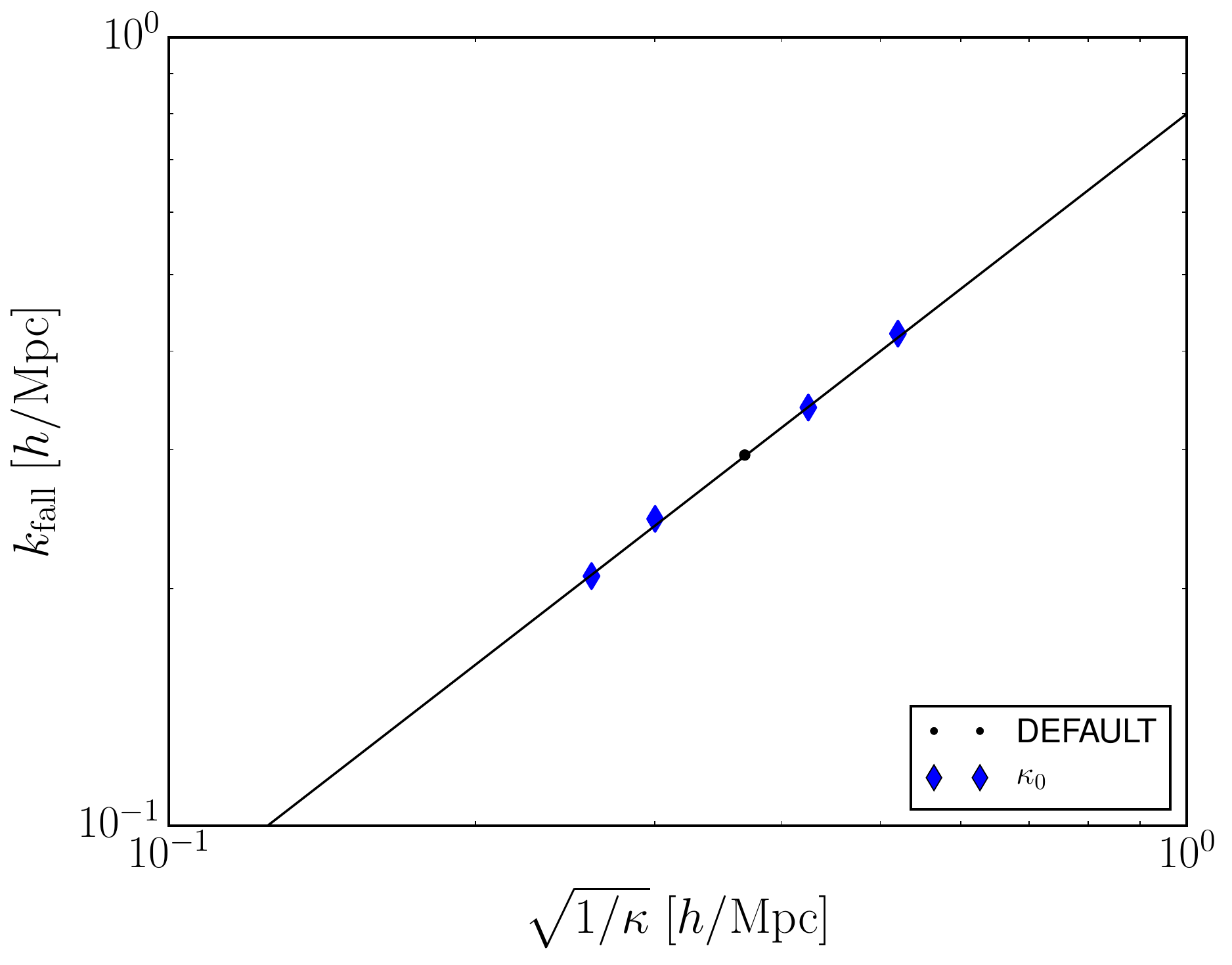}
\caption{\label{fig:fall_sys}%
\small \SP systems in the linear regime feature a characteristic length scale below which structures are strongly suppressed due to the uncertainty principle. 
We quantify this scale in Fourier space as $k_{\rm fall}$, which we found to be proportional to $\sqrt{1/\kappa}$. The diamonds correspond to a variation of $\kappa_0$ around its fiducial value (shown by the black point). The line corresponds to $k_{\rm fall}=0.8\sqrt{1/\kappa(z_{ref})}$.}
\end{figure}

In figure~\ref{fig:fall_sys}, we show the dependence of this cutoff scale on $\kappa_0$, which we find to be the single parameter that affects $k_{\rm fall}$. 
Reducing $\kappa_0$ allows structures on smaller scales to form and therefore shifts the exponential suppression to larger wavenumbers, as expected. Parametrically, we find 
\be
k_{\rm fall} \propto \frac{1}{\sqrt{\kappa}} \, ,
\ee 
which implies a slight time-dependence of this scale (in comoving momenta) of $\kappa^{-1/2} \propto \exp(\eta/4) \propto a^{1/4}$.
This confirms the interpretation as suppression related to the Jeans scale \eqref{eq:Jeans}. Note that the interpretation of the wave-function obtained from the SP equations
in terms of the Madelung representation, and the associated quantum pressure, are potentially ambiguous at these scales, as discussed above. Nevertheless, the Jeans analysis appears
to predict the correct scaling of $k_{\rm fall}$ at early redshifts.
At low redshift, additional structure on smaller scales starts to form as mentioned before.

\subsection{The amplitude problem} \label{sec:amplitude}

\begin{figure}[ht]
\centering
  \includegraphics[width=0.45\textwidth]{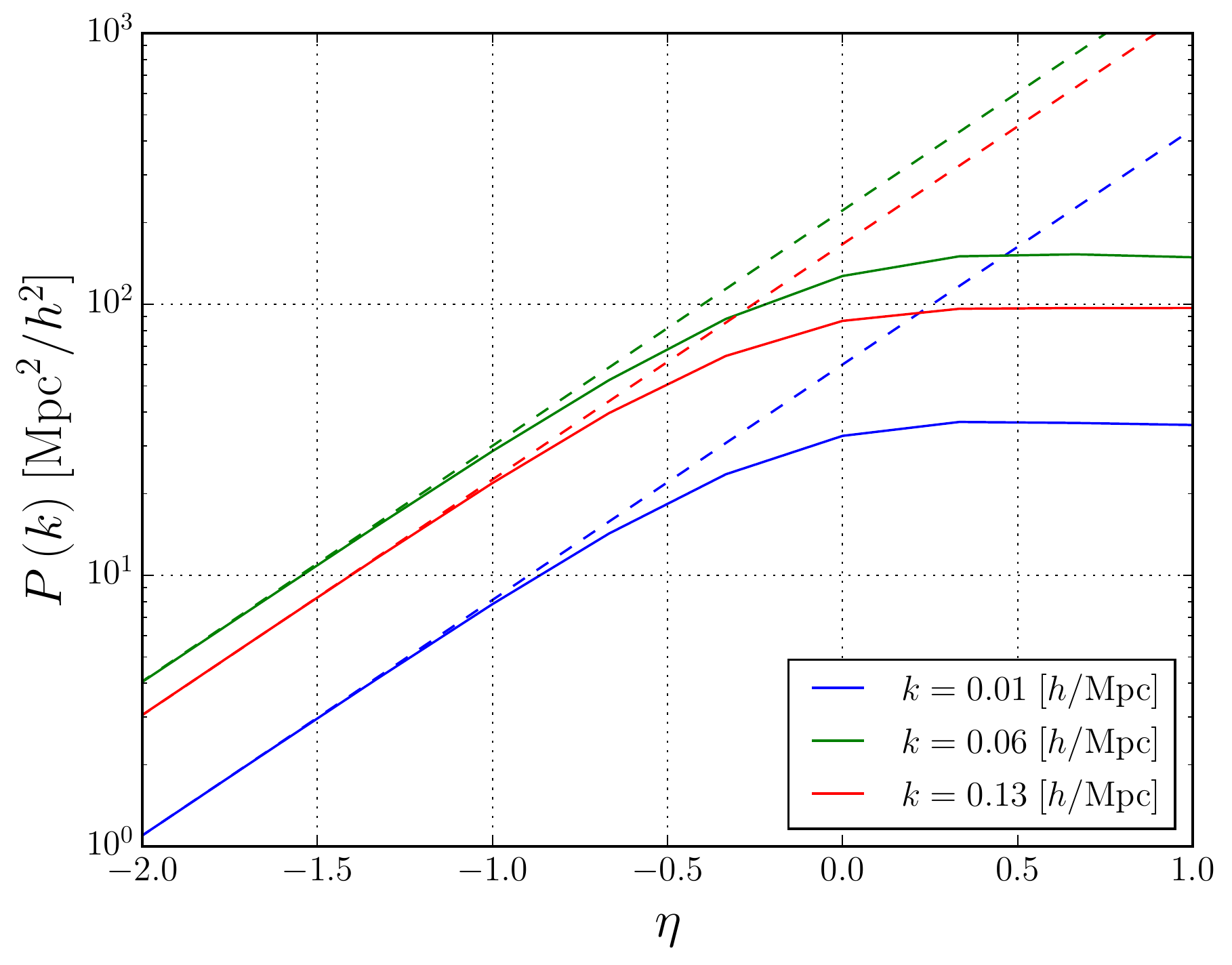}
\caption{\label{fig:amp_eta}%
\small Amplitude loss of the (2D) power spectrum versus time $\eta=\ln(a)$, for three different wavenumbers. Solid lines show the simulation result and dashed lines the expected linear growth. 
The perturbations decouple and stop growing at a particular time, which depends on the simulation parameters (see main text, the plot is for our fiducial choice). 
}
\end{figure}

The simulations show another effect that is a little bit more subtle and harder to understand. 
It is a loss of power towards the end of the simulation. 
This loss of power is evident for the smallest wavenumbers where one would expect linear evolution. In figure~\ref{fig:amp_eta}, we display the evolution of the power spectrum as a function of $\eta$ for three different modes (continuous lines). We compare with the linear evolution (dashed lines). It is possible to visualize a specific time, close to the end of the simulation ($\eta = 0$) for which each of the perturbation modes decouples and stops growing. The amplitude loss is essentially given by the amount of linear growth after this decoupling. 

\begin{figure}[ht]
\centering
  \includegraphics[width=0.45\textwidth]{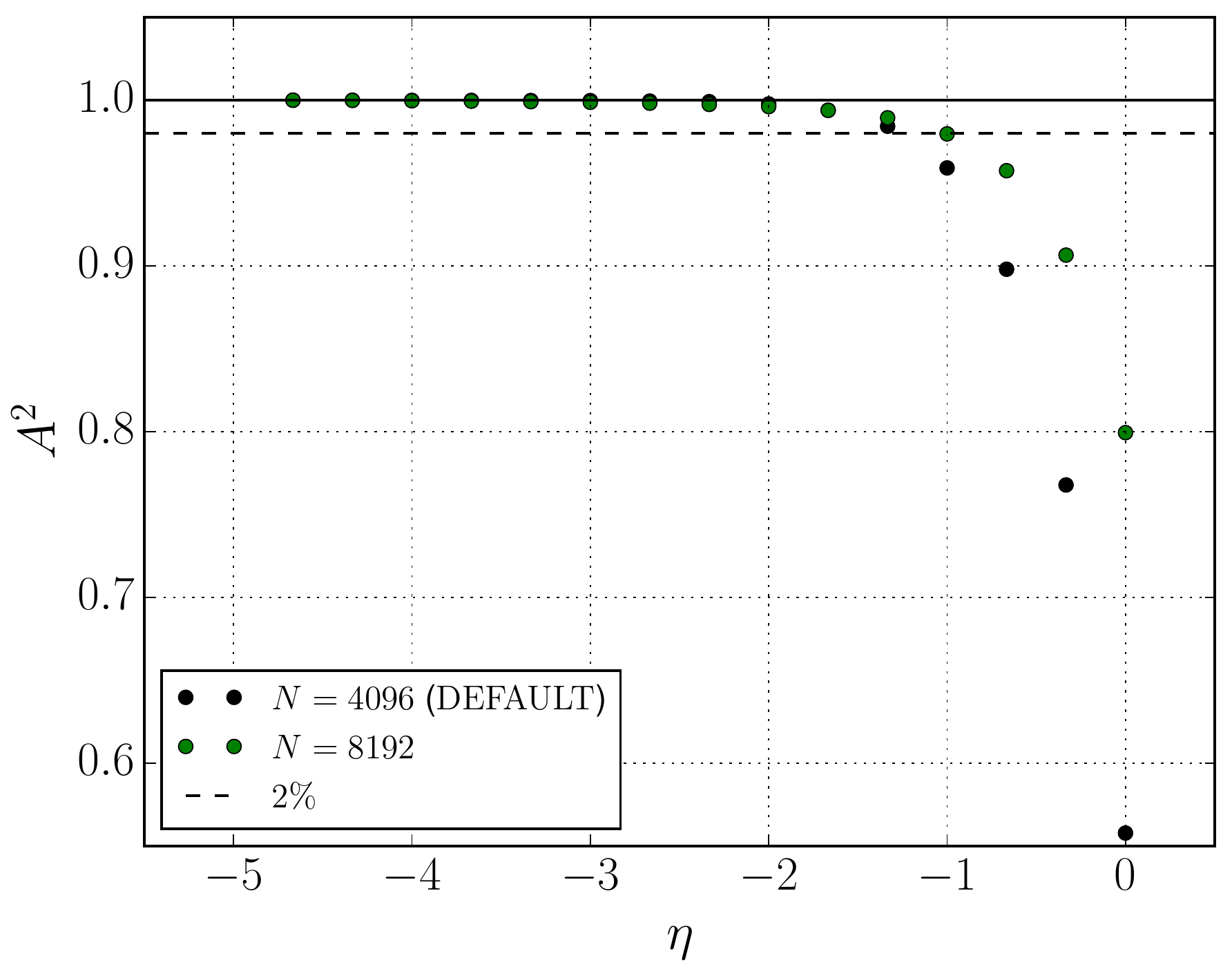}
  \includegraphics[width=0.45\textwidth]{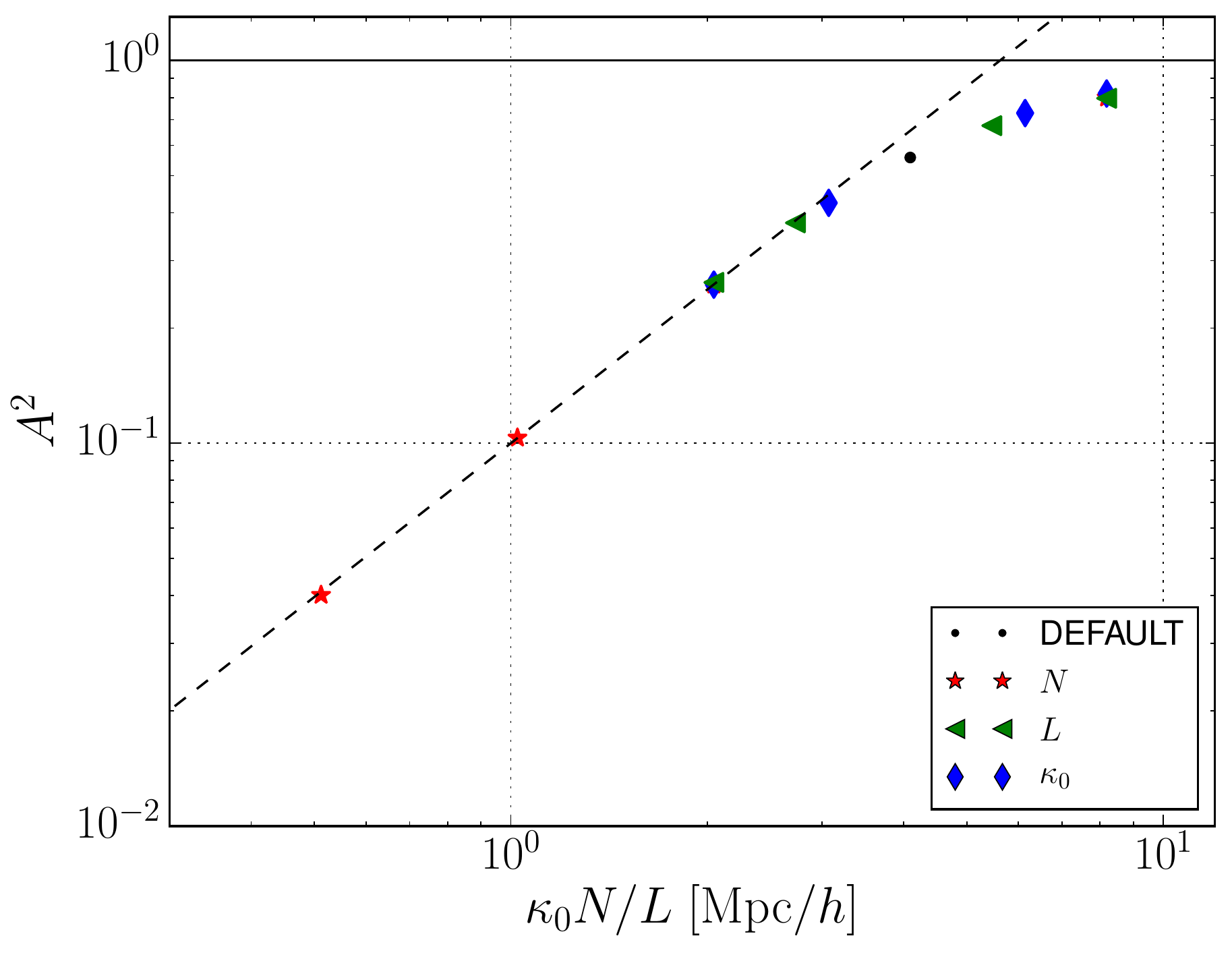}
\caption{\label{fig:amplitude_sys}%
\small Left: Amplitude loss versus time $\eta$ for the fiducial choice of parameters (black) and for a higher resolution ($N=8192$, green). Right: Dependence of the amplitude loss at $z=0$ on the combination of the simulation parameters $\kappa_0 N/L$.  The different symbols correspond to variation of either $N$, $L$ or $\kappa_0$ while keeping the other parameters fixed at their fiducial values. The dashed line corresponds to the scaling $A^2 \propto (\kappa_0 N/L)^{4/3}$,  which is inferred in the text. 
}
\end{figure}

To quantify this loss in power, we fit a straight-line coefficient $A^2$ to the ratio of the measured $P(k)$ to the rescaled power spectrum of the initial conditions $P_{\rm init}$ as expected by the linear growth function (for the modes $k<k_{\rm fall}$)
\begin{equation} \label{eq:amp}
A^2 =  \left< \frac{P}{P_{\rm init, rescaled}} \right>_{k<k_{\rm fall}}\, .
\end{equation}
In the left panel of figure~\ref{fig:amplitude_sys}, we display the evolution of $A^2$ with time $\eta$. For our fiducial set of simulation parameters, the power loss sets in at $\eta \simeq -2$ ($z \simeq 6.4$), when the non-linear evolution becomes more relevant. For larger $N$, the SP power loss is less than $2\%$ up to $\eta \simeq -1$ ($z \simeq 1.7$).
 
In the right panel of figure~\ref{fig:amplitude_sys}, we display the dependence of the amplitude loss at $z=0$ on the simulation parameters $N$, $L$, and $\kappa_0$.
We find that the amplitude loss depends only on the combination $\kappa_0 N/L$. 
It has the unit of a distance, and we find the critical length scale above which the amplitude loss effect becomes irrelevant to be 
\be
\frac{\kappa_0 N}{L} \gtrsim l_{\rm crit,0} \simeq 10 \, {\rm Mpc}/h  \, .
\ee
This can also be written as a condition on the lattice spacing $L/N$,
\be\label{eq:ampcrit}
  \frac{L}{N} \leq \frac{\kappa_0}{l_{\rm crit,0}} = \frac{\hbar}{m}\,\frac{1}{H_0 \, l_{\rm crit, 0}}\,.
\ee
In Ref.~\cite{Hui:2016ltb} it was speculated that the lattice spacing $L/N$ has to be smaller than the de Broglie wavelength $\lambda = \hbar/ m v$, where $v$ is the typical
group velocity of a wave packet. This suggests that the length scale  
$l_{\rm crit}$ is related to the velocity 
\be\label{eq:lcritdef}
 l_{\rm crit} = \frac{1}{H a} \sqrt{\langle  \mbf{u}^2 \rangle}\,,
 \ee
with $\mbf{u}$ determined from the wave-function as given by \eqref{eq:spvelocity}.
Here, we defined $l_{\rm crit}$ not at redshift zero but at general redshift as it would be 
measured on the lattice without introducing additional factors $a$ or $H$ according to (\ref{eq:spvelocity}), see discussion below.

To obtain the value of $l_{\rm crit}$ applicable to cold dark matter, one has to extrapolate the numerical results for $\langle  \mbf{u}^2 \rangle$ to $\hbar\to 0$
since $\mbf{u}$ also suffers from a suppression just as the power spectrum. We find that $l_{\rm crit, 0} \simeq 15 \, {\rm Mpc} /h$. 
Alternatively, one can estimate $l_{\rm crit}$ in linear theory. Using the linear growing mode relation $\mbf{u}=-aH\nabla\delta/\Delta$ for
the EdS background considered here gives
\be\label{eq:lcritlin}
  l_{\rm crit}^{\rm lin} =  \left(\int \frac{d^3k}{(2\pi)^3}\frac{P_{lin}(k,z)}{k^2}\right)^{1/2} \,,
\ee
which yields  $l_{\rm crit,0}^{\rm lin}\simeq 10\,{\rm Mpc}/h$. This fits well with the scale inferred from the behavior of the power loss.

\begin{figure}[ht]
\centering
  \includegraphics[width=0.45\textwidth]{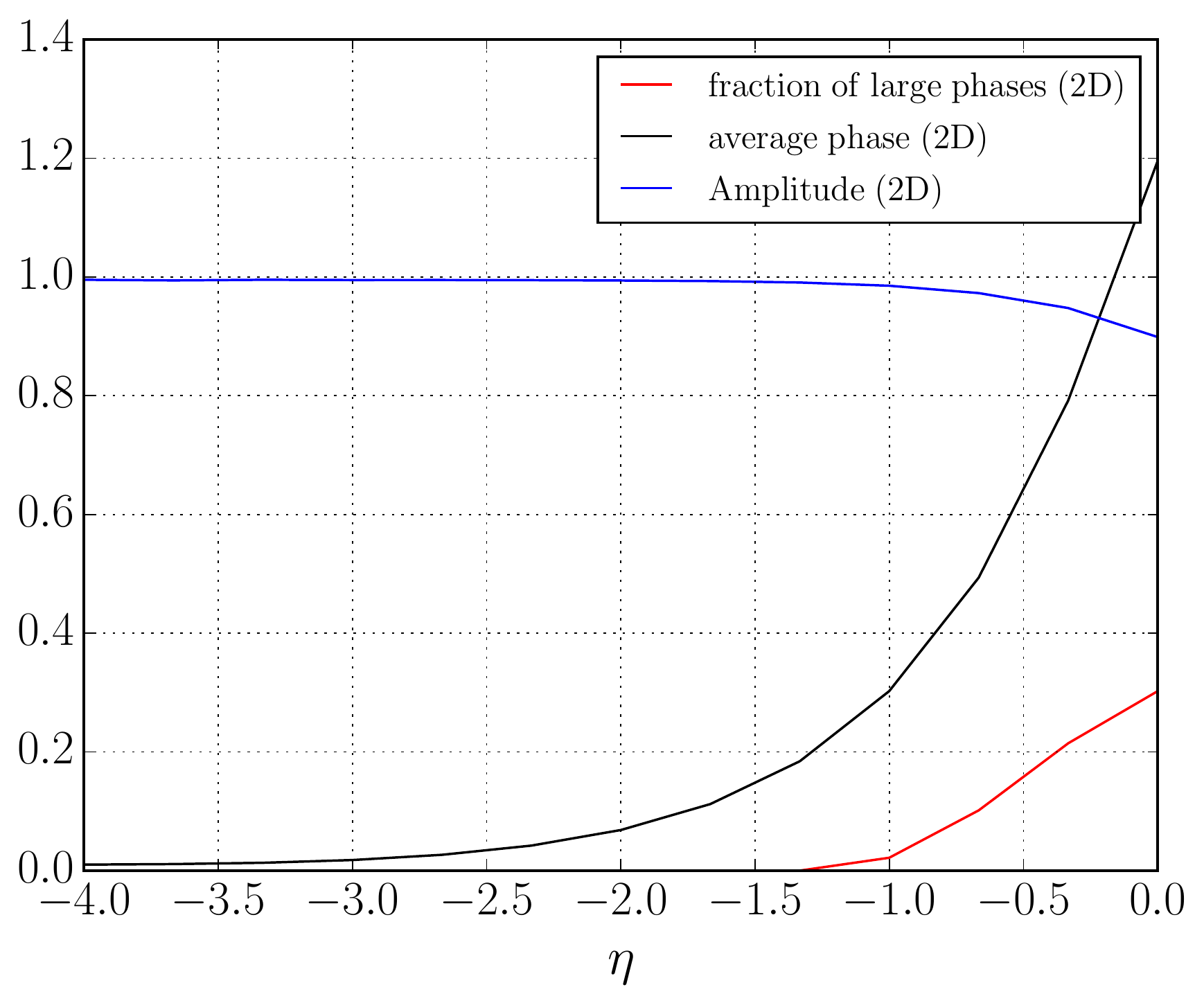}
\caption{\label{fig:2dphases}%
\small The plot shows the suppression factor $A^2$, the average of the relative phases and the fraction of large relative phases ($>\pi/4$) versus time. }
\end{figure}

Here, a couple of comments are in order.  
First, $\mbf{u}$ relates to the average peculiar velocity in the 
fluid and has no direct connection to the microscopic motion of the particles or wave packets. 
Hence, $\hbar/(m\sqrt{\langle  \mbf{u}^2 \rangle})$ strictly speaking does not represent the de Broglie wave length. 
Nevertheless, it appears to provide a valid estimate of the amplitude loss effect.
In fact, the suppression of the power spectrum probably arises from the fact that a maximal velocity exists on the lattice~\cite{Mocz:2018ium}, and is probably not intrinsic of our numerical scheme. 
Due to the spatial discretization $\nabla \phi/\hbar < \pi N /L$ which turns into the bound
\be
\frac{|\mbf{u}|}{aH} < \kappa \pi \frac{N}{L} \, .
\ee
So the relation (\ref{eq:ampcrit}) can be read as a requirement on the lattice spacing to resolve all relevant velocities in the
simulation. We tested this by studying the relative phases between neighboring grid points. 
In Fig.~\ref{fig:2dphases} we show the suppression factor $A^2$, the average of the relative phases and the fraction of large relative phases ($>\pi/4$).
The suppression happens in tandem with the occurrence of large relative phases. As a cross-check we also confirmed that energy is approximately conserved in our simulations (see Appendix~\ref{sec:econserve}) which would indicate a failure of our numerical integration of the equation of motion.

Second, note that the quantity $\langle  \mbf{u}^2 \rangle$ 
is dominated by long wavelength modes. In linear approximation, this is apparent in Fourier space, noticing that $|\mbf{u}^2| \simeq |\delta|^2(aH)^2/k^2$. 
The integral over the corresponding power spectrum in \eqref{eq:lcritlin} is dominated by modes $k\lesssim 0.1h/$Mpc.
This property fits quite well with the observation that the suppression in the power spectrum is rather wavenumber-independent.

Third, we find that at finite redshift \eqref{eq:ampcrit} is generalized to 
\be
\frac{L}{N} \leq \frac{\kappa}{l_{\rm crit}} = \frac{\hbar}{m}\,\frac{1}{a^2 \, H \, l_{\rm crit}}
\simeq \frac{\hbar}{a\, m \sqrt{\langle  \mbf{u}^2 \rangle}}\,, 
\ee
Notice that the time-dependence in (\ref{eq:lcritlin}) implies $l_{\rm crit} \propto a$ in the linear regime and 
hence $\kappa/l_{\rm crit} \propto a^{-3/2}$. Therefore, the amplitude loss sets in when
\be\label{eq:acritdef}
  a \geq a_{\rm crit} \simeq \left(\frac{\kappa_0N/L}{l_{\rm crit,0}}\right)^{2/3}\,.
\ee
For the simulation parameters shown in the left panel of figure~\ref{fig:amplitude_sys}, this corresponds to
$\eta_{\rm crit}=\ln a_{\rm crit}\simeq -0.4\, (-0.94)$ for $N=8192\, (4096)$, in good agreement with our numerical findings.
The observation that the power spectrum essentially saturates after the power loss sets in leads to another prediction.
Because $A^2$ is dominantly determined by large scales, for which the conventional linear power spectrum grows as $a^2$, 
\eqref{eq:acritdef} implies that 
\be
  A^2 \propto (\kappa_0 N/L)^{4/3}
\ee
within the regime $a_{\rm crit}\ll 1$, and $A^2\to 1$ for $a_{\rm crit}\gg 1$. This expectation is confirmed
by our numerical results shown in the right panel of figure~\ref{fig:amplitude_sys}.

Finally, we checked that the amplitude loss stems only from the spatial discretization. 
Changing the time-like discretization has no impact on the effect, as seen in appendix~\ref{app_convergence}.

\subsection{The (sampling) noise problem} \label{sec:noise}

\begin{figure}[ht]
\centering
  \includegraphics[width=0.45\textwidth]{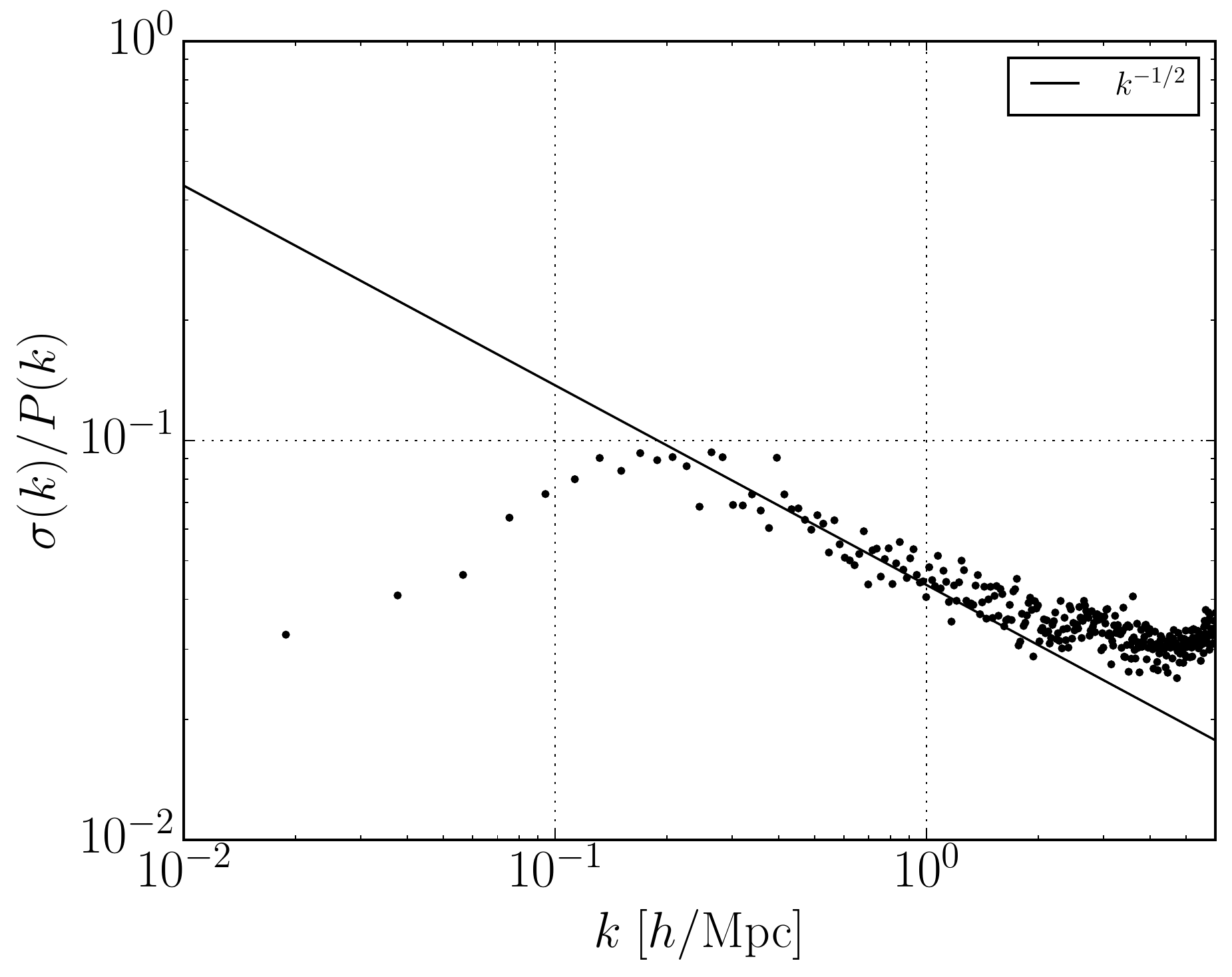}
\caption{\label{fig:noise_sys}%
\small Sampling variance of the power spectrum at $z=0$ obtained from 64 realizations, normalized to the power spectrum as a function of the wavenumber. We also show a line
that corresponds to the expected scaling based on the number of Fourier modes in 2D.}
\end{figure}

The initialization of the wave-function that is used in this work features random phases for each Fourier mode, while the amplitude is fixed (see section~\ref{sec:ic}).
The absence of random fluctuations in the initial amplitude tends to decrease sampling variance.
Nevertheless, due to the non-linear dynamics, the measured power spectrum is affected by (sampling) noise resulting from the initial random phases as set up in (\ref{eq:Ps_initial}). 
As expected, in the power spectrum this noise becomes smaller with $k$ due to the growing number of Fourier modes contained in the simulation volume.
This is seen in figure~\ref{fig:noise_sys}, where we display the variance measured from 64 simulations with different initial seeds (and otherwise fiducial parameters).
For very small wavenumbers, the noise is suppressed due to our choice of initial conditions, and since the dynamics is almost linear. 
For larger wavenumbers, one expects that the variance of the power spectrum normalized by $P(k)$ scales as $1/\sqrt{k}$ in 2D and $1/k$ in 3D, which is
reproduced in our simulations. For very large wavenumbers, the variance further decreases, but a drop in the power spectrum leads to the increasing noise in figure~\ref{fig:noise_sys} due to the normalization chosen. We could not identify any `shot noise' in the simulations in the sense of a wavenumber-independent noise component.

For a fixed comoving momentum, the noise can be reduced by increasing the simulated volume, since this increases the number of modes that represent this momentum in Fourier space by $(kL/2\pi)^{D-1}$. 
The sampling noise may also be further suppressed by using the technique of paired initial phases proposed in \cite{Angulo:2016hjd} for $N$-body simulations. In 3D the noise is always strongly reduced due to the different scaling $1/k$.

\subsection{Summary of systematic effects}

When using the SP method to describe cold dark matter, one would, in principle,
like to take the limit $\kappa_0 \propto \hbar \rightarrow 0$ the same way one would want to set the particle mass as small as possible in $N$-body simulations. 
The price to pay when decreasing $\kappa_0$ is twofold: First, the computational time increases because the argument of potential rotations $U_V$ becomes larger, which
requires to reduce the time step [see (\ref{eq:short_definitions})]. Second, lowering $\kappa_0$ makes the amplitude loss problem described above more severe. 
The best alternative would then be to reduce $\kappa_0$ while increasing $N$, at the cost of more demanding simulations. 

In order to mitigate the loss of power at late times, one can either increase $\kappa_0$ or make the lattice spacing $L/N$ smaller. 
The first alternative comes with the price of an exponential suppression at a smaller $k_{\rm fall}$.
Increasing $N$ increases the computational cost (see appendix~\ref{app:time}), while decreasing $L$ increases the sampling noise. 
For the 2D simulation with $L=1000$~Mpc$/h$, $\kappa_0 = 1$ Mpc$^2/h^2$ and $N = 8192$, we measured the amplitude loss, Jeans suppression and sampling noise at $z=0$, 
to be
\bea
A^2 &=& 0.8 \,, \\
k_{\rm fall} &=& 0.3 \, h/{\rm Mpc}\,,\\
\sigma/P &\sim & 10\%\, .
\eea 
In this context, it is interesting to explore the 1D case, for which we can substantially increase the resolution (see appendix~\ref{app:1d}). 
In that case, we can decrease the loss in power for the SP system for $N = 80.192$  down to the percent level.

Instead of increasing $\kappa_0N/L$, one may wonder whether it is possible to apply a correction that compensates for the amplitude power loss.
The simplest possibility is to rescale the power spectrum by $1/A^2$. However, the extent to which this naive rescaling captures non-linear growth is
unclear. Nevertheless, we followed this approach to investigate the correlation function around the BAO peak (see below). 
Alternatively, one could run the simulation somewhat longer in the hope that this captures the non-linear effects better than just a rescaling. 
However, it turns out that this only works poorly since the growth rate of the power spectrum features a plateau once the power loss sets in, see figure~\ref{fig:amp_eta}. 

\begin{figure}[ht]
\centering
  \includegraphics[width=0.45\textwidth]{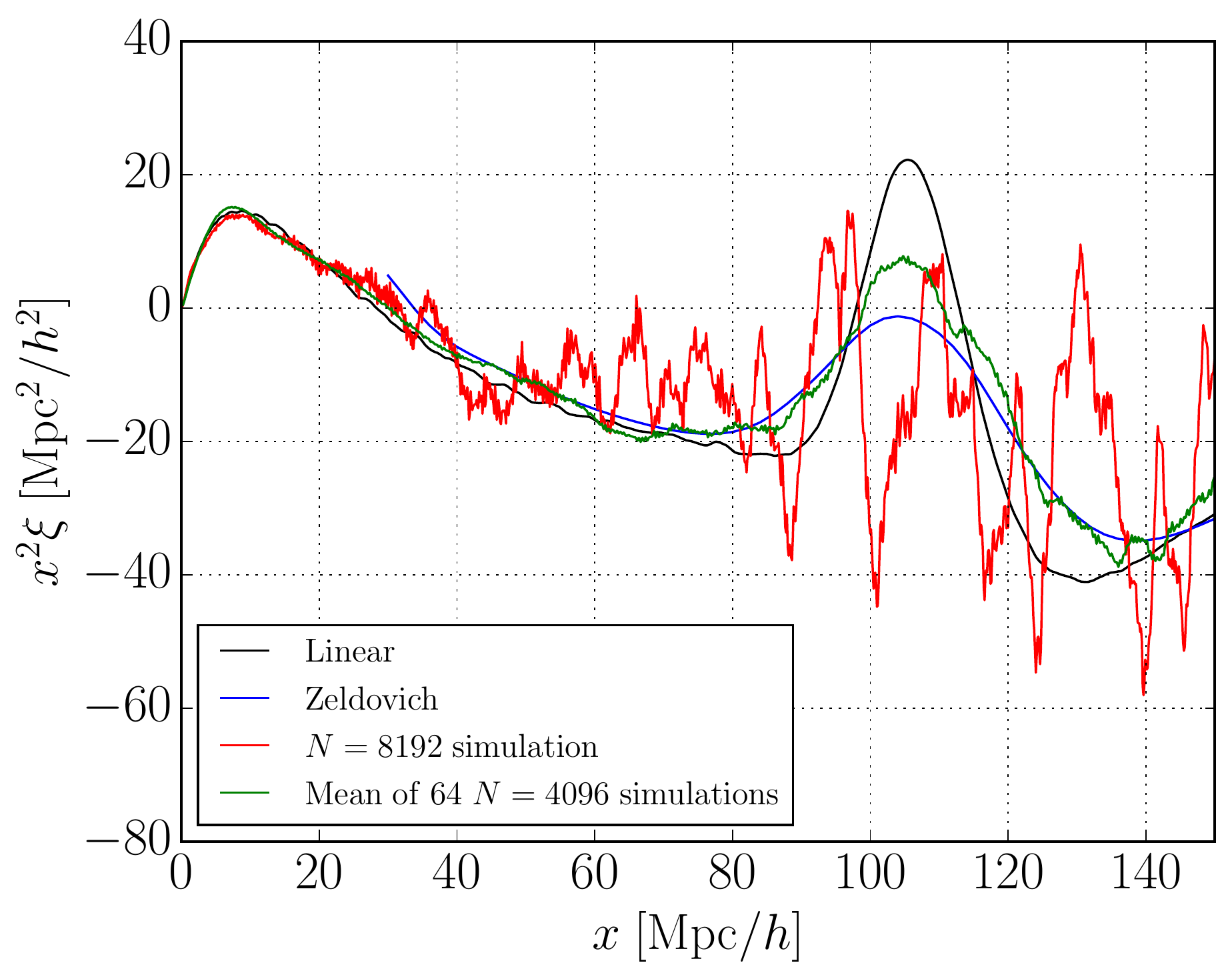}
  \includegraphics[width=0.45\textwidth]{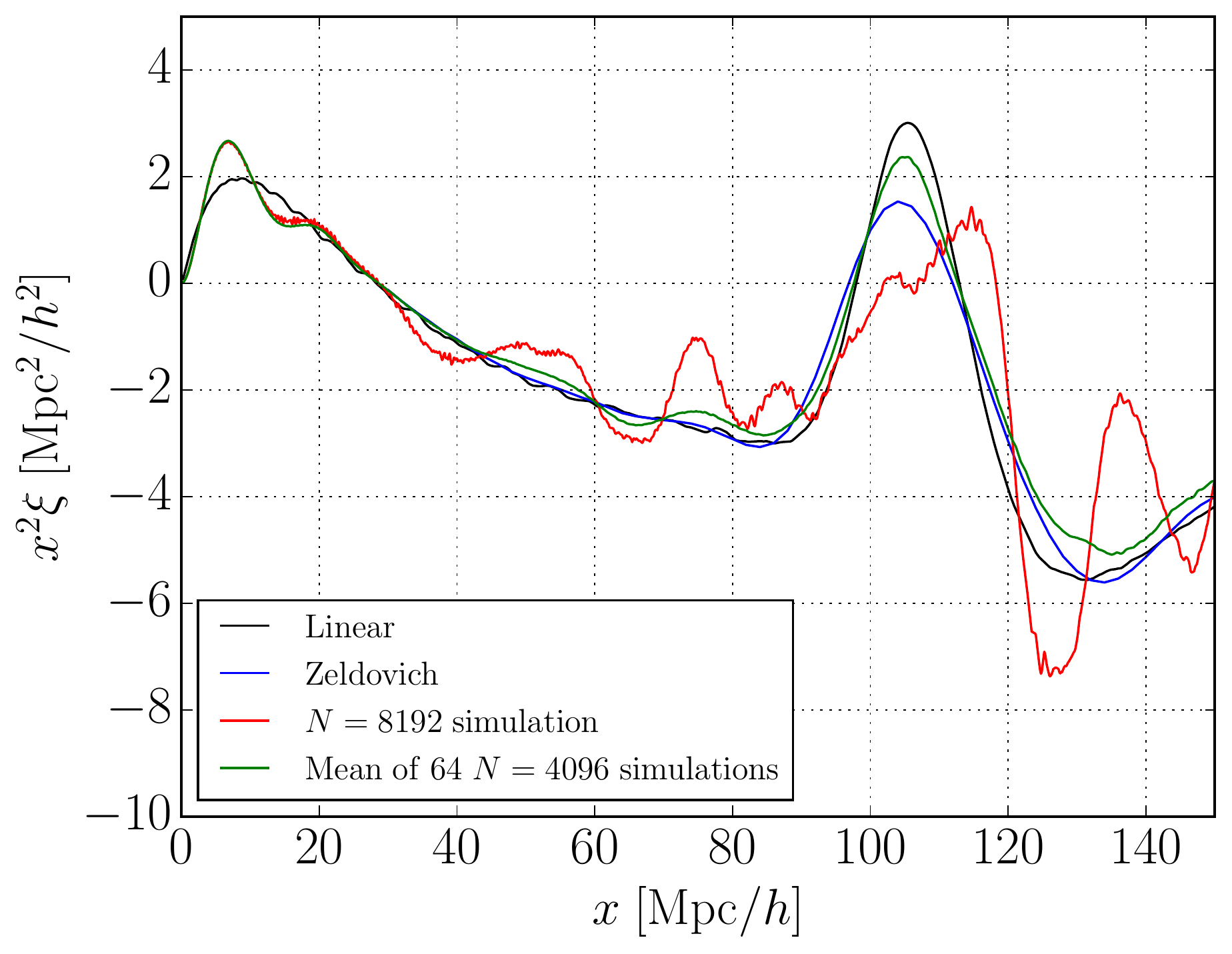}
\caption{\label{fig:composing_smallboxes}%
\small Correlation function $x^2\xi(x)$ for redshift $z=0$ (left) and $z=1.72$ (right). We show the SP result obtained from averaging over 64 simulations with fiducial parameters, from a single simulation
with parameters as in figure~\ref{fig:pdf_z_2d}, in linear theory and for Zel'dovich approximation.   }
\end{figure}

Another naive approach would be to compensate the loss of power by a rescaling the power spectrum. By construction, this would lead to the correct overall normalization of the 
power spectrum but will also fail once non-linear features are relevant.
In figure~\ref{fig:composing_smallboxes},
we display the correlation function measured at two redshifts, which is obtained by averaging over 64 simulations (with $N = 4096$). 
We also display the result of a single simulation (with $N = 8192$) for comparison, as well as the linear correlation function
and the  prediction in Zel'dovich approximation. Here we rescaled the correlation function obtained from SP 
by $1/A^2$, where $A^2$ is the redshift-dependent power loss determined in section~\ref{sec:amplitude}. 
The origin of the noise in the correlation function can partially be attributed to sampling variance, and partially
to fluctuations of the SP system at smaller wavenumber. The averaging over 64 simulations reduces the noise considerably, as expected, such
that the BAO peak becomes visible. Notice that the correlation function at larger redshift appears to be less noisy, which is due to the 
suppression of small scale fluctuations by the Jeans scale (c.f.~the power spectrum in figure~\ref{fig:pdf_z_2d}).

The corresponding correlation function is close to the Zel'dovich approximation at both redshifts, except in the vicinity of the
BAO peak. The broadening of the BAO peak is less pronounced compared to Zel'dovich approximation.
Several systematics could induce the lack of BAO broadening: Apart from the dynamical range that is
limited by the box size and the Jeans suppression scale at small and large wavenumber, also the amplitude power loss could play a role. 
The latter effectively shuts off the growth of perturbations on (and above) BAO scales,
leading to a lack of non-linear BAO damping that cannot be compensated by linear rescaling of the amplitude. We quantitatively checked this by calculating the 
Zel'dovich approximation including the power loss but found that it does not seem to be the main reason for the lack of BAO broadening.
On the other hand, the existence of a maximal velocity in the simulation qualitatively explains the lack of broadening in the correlation function as well as the lack of power in the power spectrum (see section \ref{sec:amplitude}). Still, there seems to be no simple way to counteract this effect and simulations 
with finer grids seem indispensable.

\section{Towards 3D in the \SP method}
\label{sec:3d_sp}

\begin{figure}[ht]
\centering
  \includegraphics[width=0.45\textwidth]{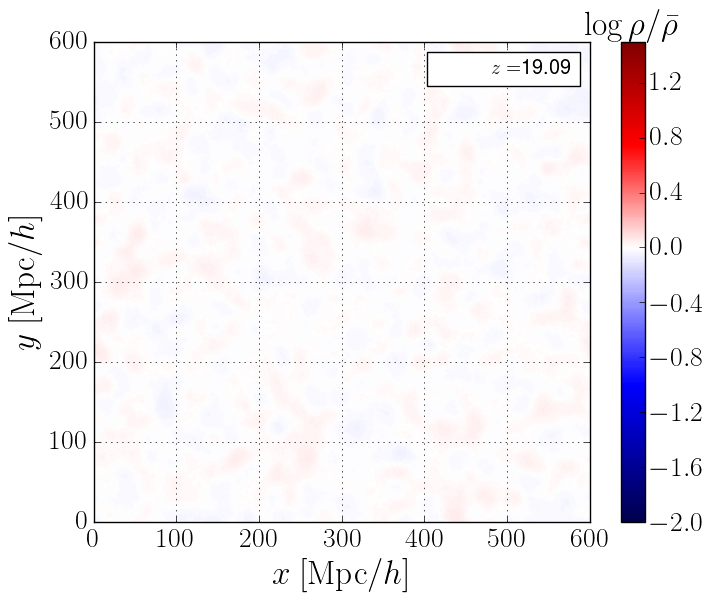}
  \includegraphics[width=0.45\textwidth]{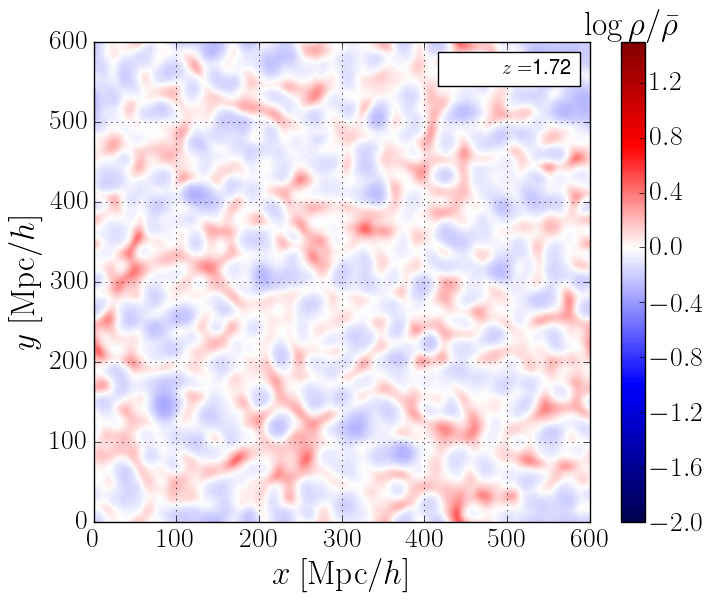}  
  \includegraphics[width=0.45\textwidth]{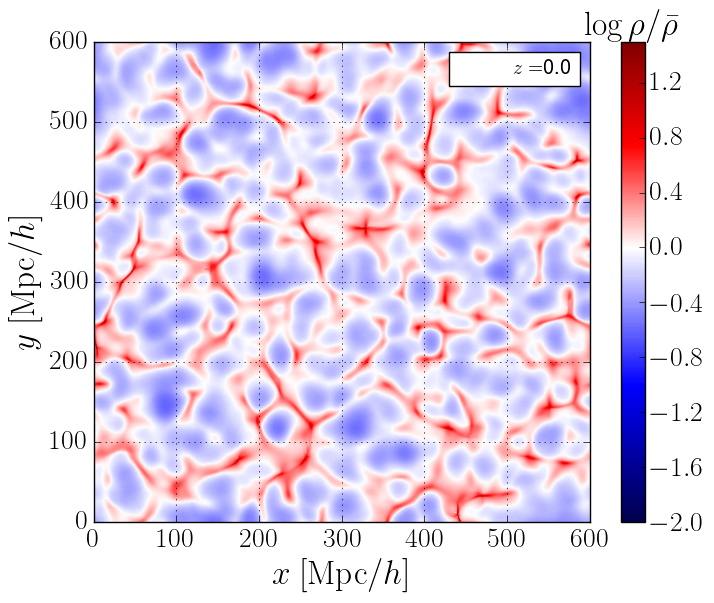}
  \includegraphics[width=0.45\textwidth]{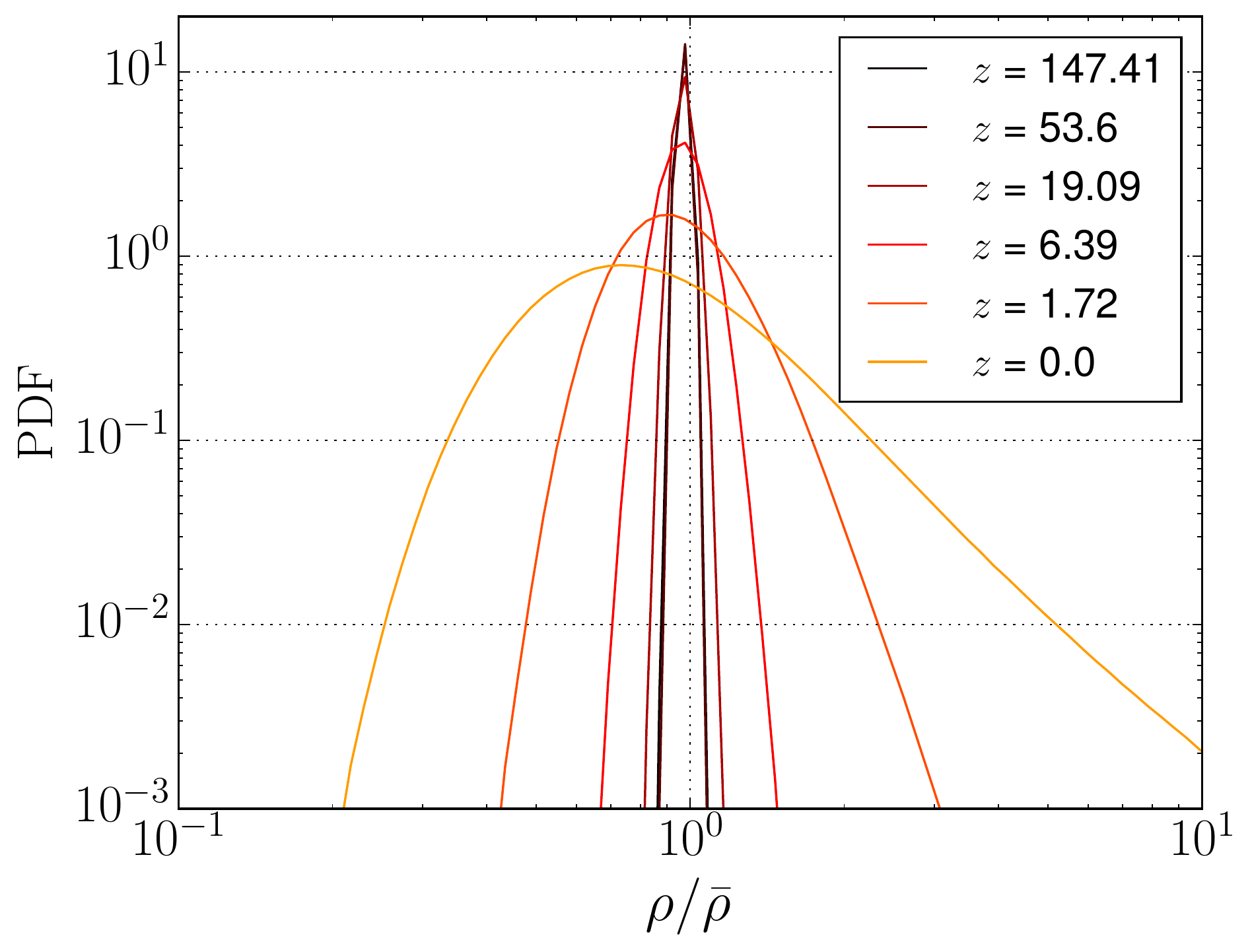}  
\caption{\label{fig:dens_3d_z}%
\small Density field at three different redshifts ($z=19.09$, $z=1.72$ and $z=0$) for the 3D \SP system. We projected the density field taking the mean of 10 slices. In the bottom right panel, the PDF of the density field at various redshifts $z$ is shown.}
\end{figure}

In this section, we present solutions of the \SP system in 3D. 
The main motivation is not to present any physical results -- which will be numerically even more demanding than in 2D. The purpose 
of this section is to validate if the three systematics identified in 2D show the same parametric dependence and to estimate 
the scale $l_{\rm crit}$.
We choose simulation parameters $L=600$ Mpc$/h$, $N = 512$ and $\kappa_0 = 4 \,{\rm Mpc}^2/h^2$. 
In figure~\ref{fig:dens_3d_z}, we display the overdensity field for three different redshifts 
in a slice of the simulation volume, after calculating the mean of 10 bins along the $z$ axis. 

In the bottom right panel of figure~\ref{fig:dens_3d_z}, the PDF of the density field for different redshifts is shown. Even though the PDF departs from its initial shape, developing some skewness and kurtosis, it is still far from developing the non-linear shape found in 2D (see e.g.~figure~\ref{fig:pdf_z_2d}).

\begin{figure}[ht]
\centering
  \includegraphics[width=0.45\textwidth]{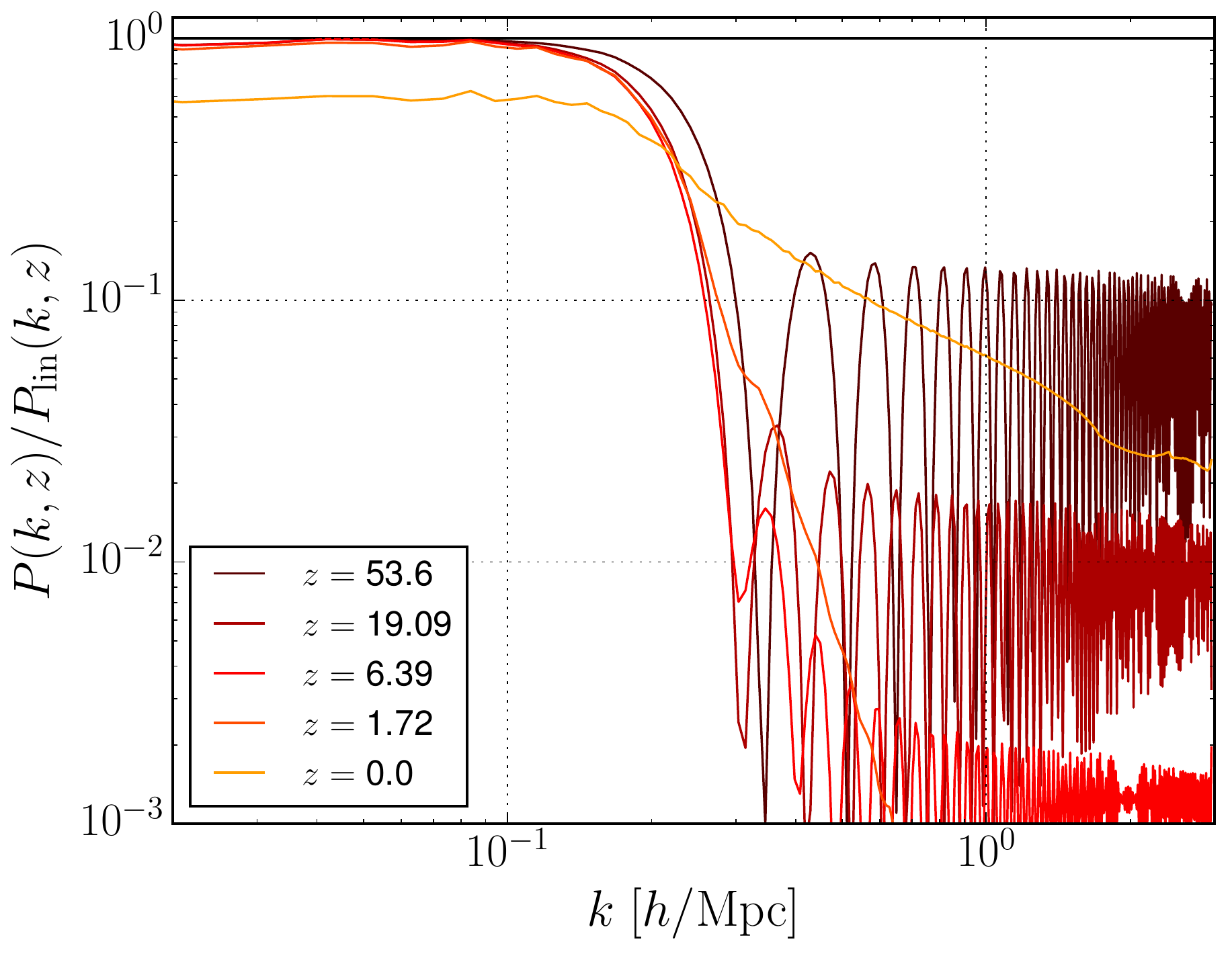}
\caption{\label{fig:pk_3d}%
\small Power spectrum obtained from a 3D SP simulation at various redshifts $z$, using $L=600$ Mpc $/h$, $N = 512$ and $\kappa_0 = 4 \,{\rm Mpc}^2/h^2$.}
\end{figure}

\begin{figure}[ht]
\centering
  \includegraphics[width=0.45\textwidth]{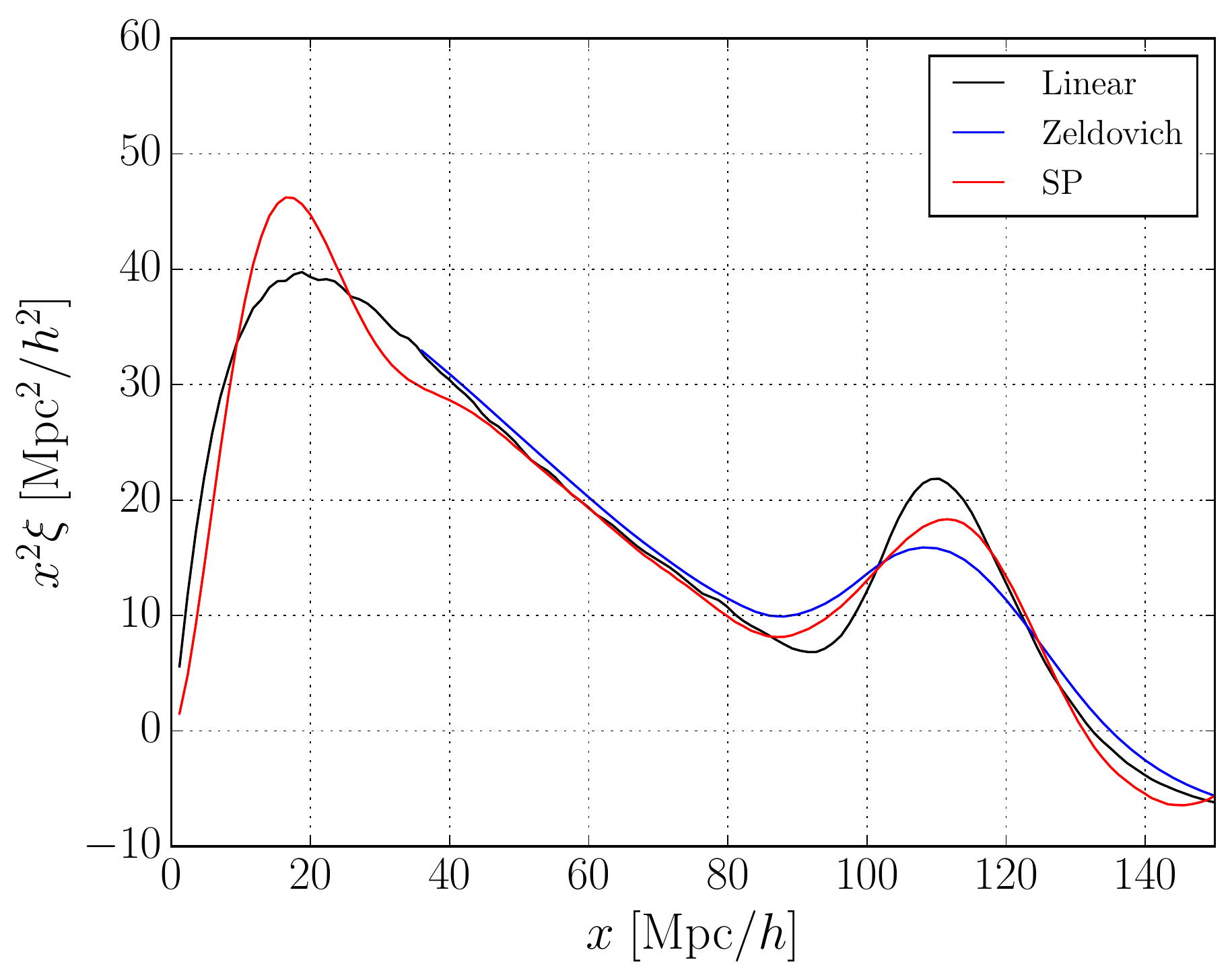}  
  \includegraphics[width=0.45\textwidth]{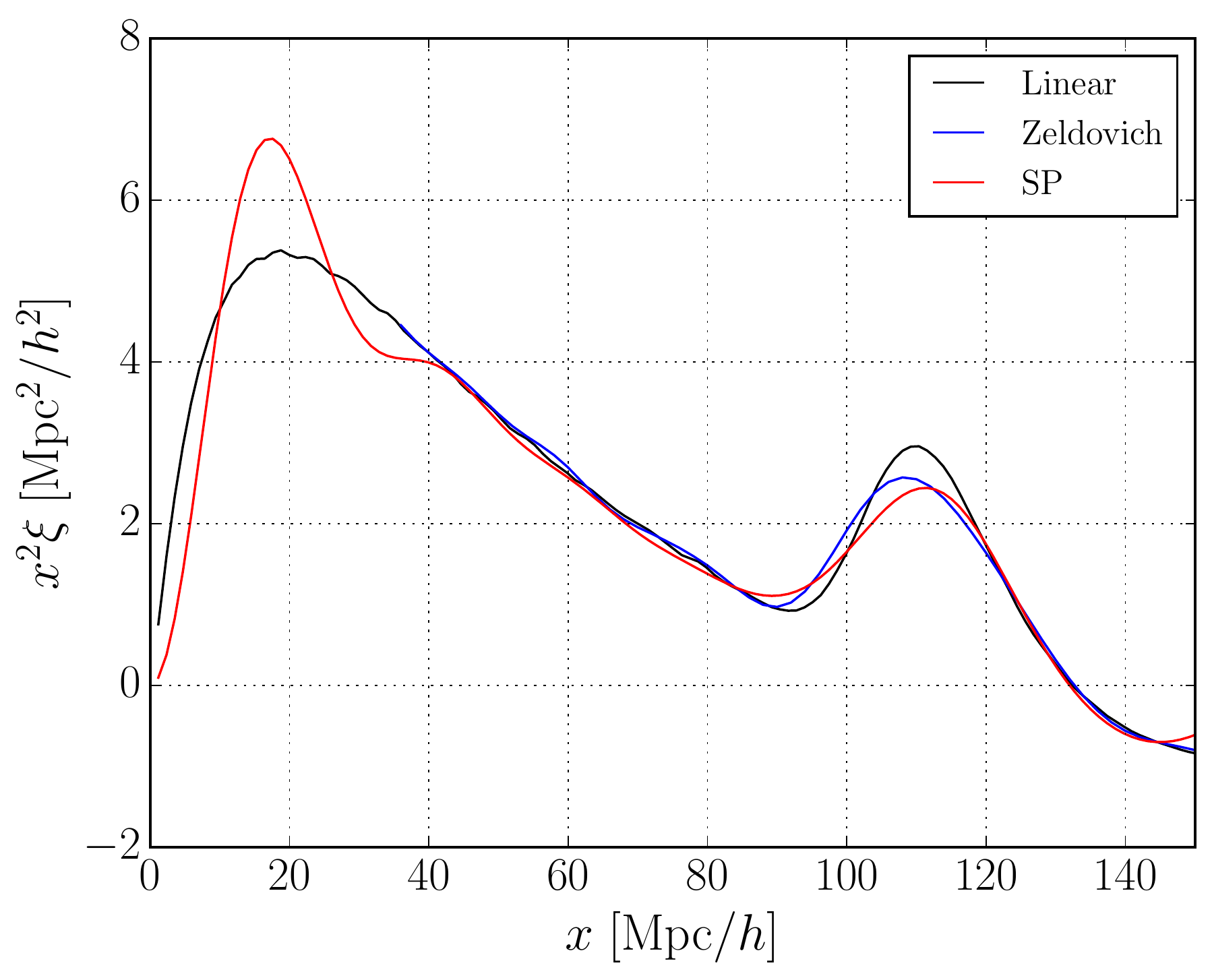}  
\caption{\label{fig:corr_3d}%
\small Correlation function for the 3D SP simulation at $z =0$ (left) and $z = 1.27$ (right).}
\end{figure}

The power spectrum is shown in figure~\ref{fig:pk_3d}. 
It features Jeans-like suppression at large $k$ as in 2D, as well as an overall amplitude loss at low redshift.
For the 3D simulation, the parameters characterizing the overall power loss and the Jean suppression scale defined in section~\ref{sec:systematics} are found to be (at $z=0$)
\bea
A^2 &=& 0.6 \,, \\
k_{\rm fall} &=& 0.15 \, h/{\rm Mpc}\,.
\eea
Notice that the power loss in terms of $A^2$ is in accordance with the parametric 
dependence on simulation parameters identified in the 2D case. 
In particular, for the 3D simulation $\kappa_0 N/L \simeq 3.4$ Mpc$/h$, which implies that $l_{\rm crit}$ is close to the 2D value.

In order to obtain acceptable values of the power loss, the box size has been reduced and $\kappa_0$ increased compared to 2D. The latter leads to a smaller $k_{\rm fall}$. 
In principle, one could increase $k_{\rm fall}$ while keeping $A^2$ fixed by decreasing $\kappa_0$ and $L$. 
However, this is not possible since the BAO peak has to fit into the box. 
Ultimately, one will have to keep the box size fixed and increase $N$.
The sampling noise is substantially reduced compared with the 2D case, because the number of modes for a fixed momentum $|k|$ is larger and scales as $(k\,L)^2$. 

The correlation function (see figure \ref{fig:corr_3d}) extracted from a single realization is substantially less affected by noise as compared to 2D.
As before, we rescaled $\xi(x)$ extracted from the simulation by $1/A^2$ at each redshift. The result is then found to be relatively close to
the Zel'dovich approximation for $z=1.72$, while a slight lack of BAO broadening is visible at $z=0$, similar to the 2D case.

\section{Conclusion} \label{sec:conclusion}

We studied the growth of large-scale structure at BAO scales using the \SP approach for cold dark matter. The main question is if large-scale simulations, competitive with $N$-body simulations, are feasible in this setup. The appeal of a second independent approach to large-scale structure is that the \SP method comes with a different methodology for initial conditions, dynamics, no gravitational softening and hence different systematic uncertainties. Besides, it makes higher moments of the phase space distribution function and velocity correlation functions more readily available. 
We identified three systematic effects (for most parts already seen previously in refs.~\cite{Khlopov:1985jw, Hu:2000ke, Li:2018kyk}) and studied their parametric dependence on the simulation parameters. There is a Jeans damping scale, an overall suppression of the amplitude (due to a lack of resolution of the wave packets) as well as sampling noise. We provide a quantitative criterion to determine the redshift after which amplitude suppression sets in, and find a particular combination of simulation parameters it depends on. In order to avoid this effect, the simulation parameters should obey
\be
\label{eq:deBroglie}
 \frac{L}{N} \leq \frac{\kappa}{l_{\rm crit}} \simeq \frac{\hbar}{a\, m \sqrt{\langle  \mbf{u}^2 \rangle}}\,.
\ee
We interpret this criterion in terms of an effective de Broglie wavelength and the existence of a maximal velocity in the simulation. 

\begin{figure}[ht]
\centering
  \includegraphics[width=0.9\textwidth]{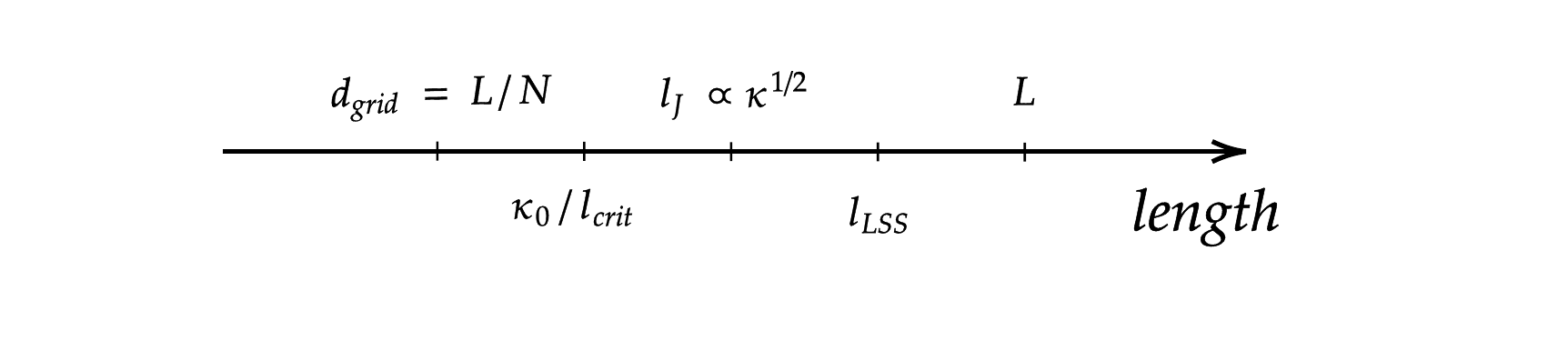}
\caption{\label{fig:diagram}%
\small Schematic illustration of all the length scales occurring in the \SP simulation. In order to control the exponential fall, the Jeans suppression scale $l_{J}$ ought to be smaller than the BAO scale $l_{LSS}$. The noise is suppressed when the size of the box $L$ is much larger than the BAO scale. To overcome the overall amplitude loss, the grid size $d_{\rm grid}$ must be smaller than $\kappa_0/l_{\rm crit}$ (see section~\ref{sec:amplitude} and (\ref{eq:deBroglie})).}
\end{figure}

The main challenge in 3D is to clearly separate all the occurring scales in the simulations (see figure~\ref{fig:diagram}).
Ideally, the Jeans scale should be substantially smaller than the BAO scale and the box size substantially bigger. Furthermore, the grid spacing should be substantially smaller than the effective de Broglie wavelength, see (\ref{eq:deBroglie}). All in all, this requires large grid sizes for accurate simulations ($N> 16$k in all dimensions) and the simulations are rather memory bound than compute bound. Overall, we find it realistic that the \SP simulations for cold dark matter clustering could become competitive with $N$-body simulations. The algorithm used in the present analysis is very basic and hopefully more sophisticated techniques will be developed in the future and tap into the true potential of the \SP method.

\section*{Acknowledgments}
We thank Oliver Hahn, Lam Hui, Simon May and Cora Uhlemann for discussions. TK and HR are funded by the Deutsche Forschungsgemeinschaft under Germany‘s Excellence Strategy – EXC 2121 "Quantum Universe" – 390833306. This research was supported by the German Research Foundation cluster of excellence
ORIGINS (EXC 2094, \href{www.origins-cluster.de}{www.origins-cluster.de}).
\appendix

\section{Convergence test}\label{app_convergence}

As explained in section~\ref{sec:sp_algo}, to ensure the convergence of the wave function, we define a maximum angle $\theta_{\rm max}$ for the rotations $U_K$ and $U_V$ in equation~(\ref{eq:rotations}). If either the rotation angle of the potential or the kinetic part is higher than this value, we reduce the time step $\Delta s$. In case one of the angles for one of the modes is too large, a sizable error  accumulates quickly.  

\begin{figure}[ht]
\centering
  \includegraphics[width=0.45\textwidth]{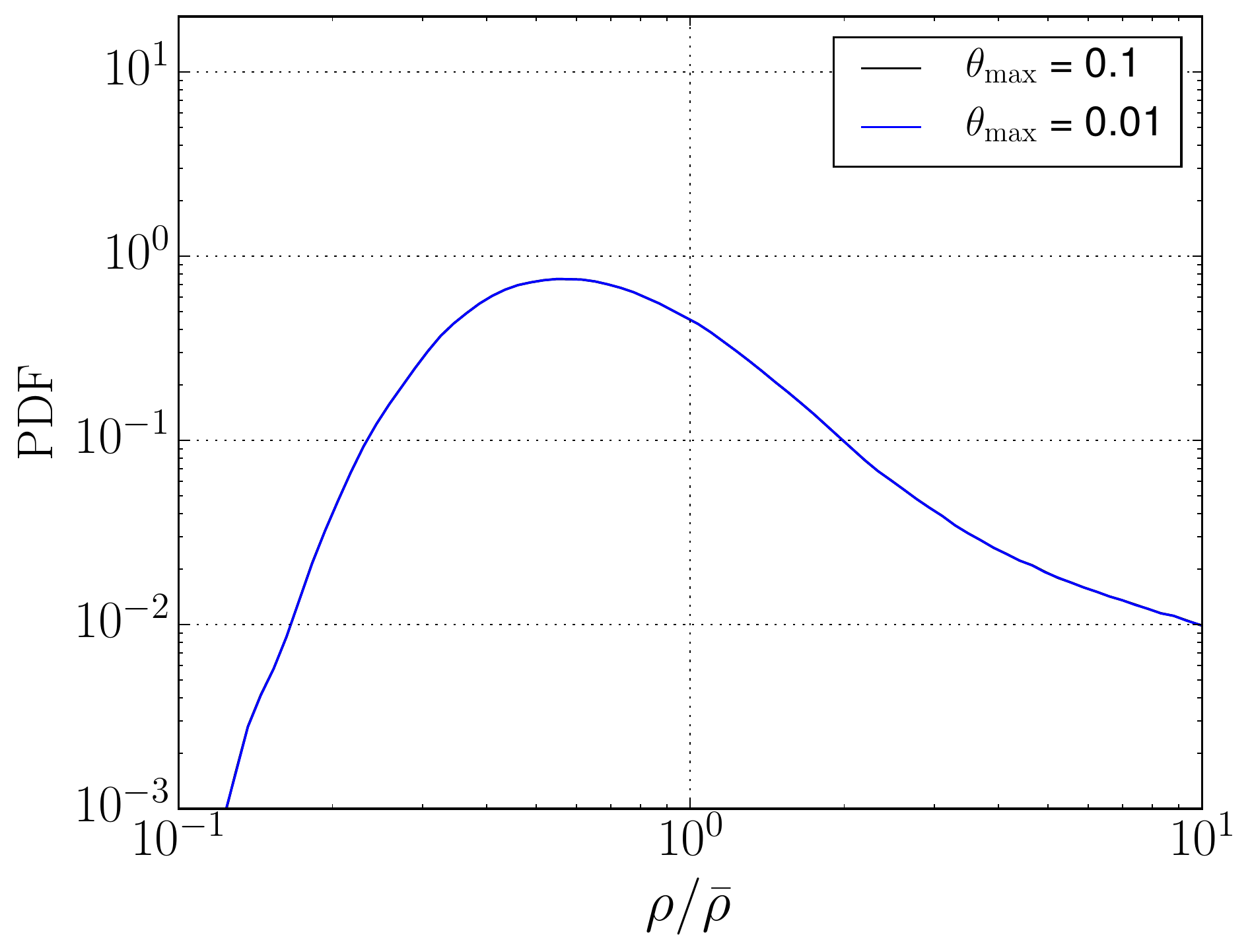}  
  \includegraphics[width=0.45\textwidth]{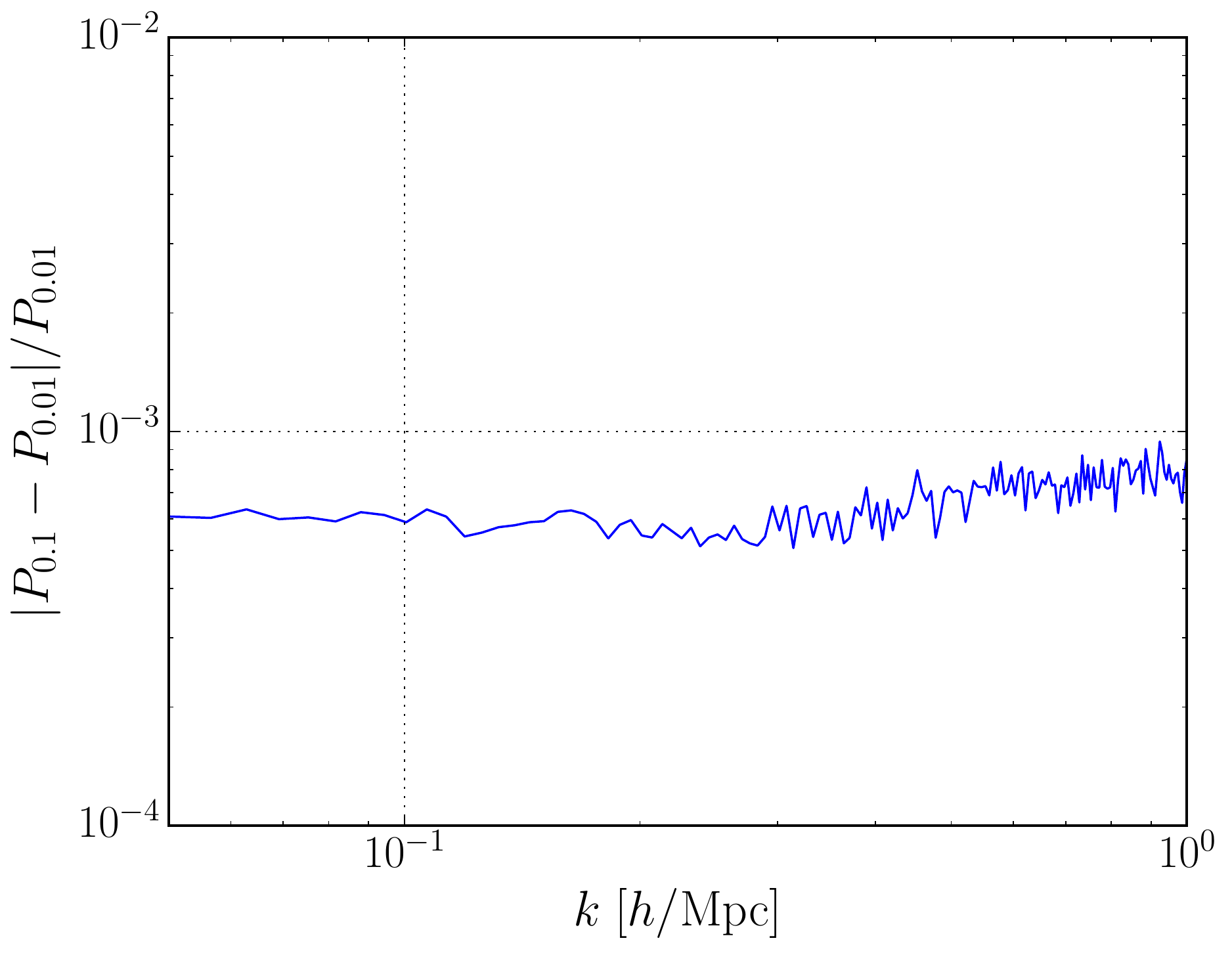}
\caption{\label{fig:convergence}%
	\small Convergence of the SP solution when reducing the time step. In the left, the PDF obtained for $\theta_{\rm max} = 0.01$ and $\theta_{\rm max} = 0.1$ is shown. The PDFs overlap, being indistinguishable. In the right, the relative difference in the power spectrum is shown.}
\end{figure}

In figure~\ref{fig:convergence}, we show the effect of reducing the value of \textbf{$\theta_{\rm max}$} -- and therefore increasing the computational time -- in the overdensity distribution (left) and in the power spectrum (right). In this work we used $\theta_{\rm max} = 0.1$ and here we compare with using $\theta_{\rm max} = 0.01$ instead. In the left panel, both PDFs overlap. In the right panel, we show the relative difference of the density power spectrum, which is below $10^{-3}$ over the entire range of scales considered in this work. The relative difference of the wave-function $\psi$ is of the order of $\sim 10^{-6}$. We conclude that using $\theta_{\rm max} = 0.1$ is sufficient to guarantee numerical stability. 

\section{Energy conservation\label{sec:econserve}}

\begin{figure}[ht]
\centering
  \includegraphics[width=0.45\textwidth]{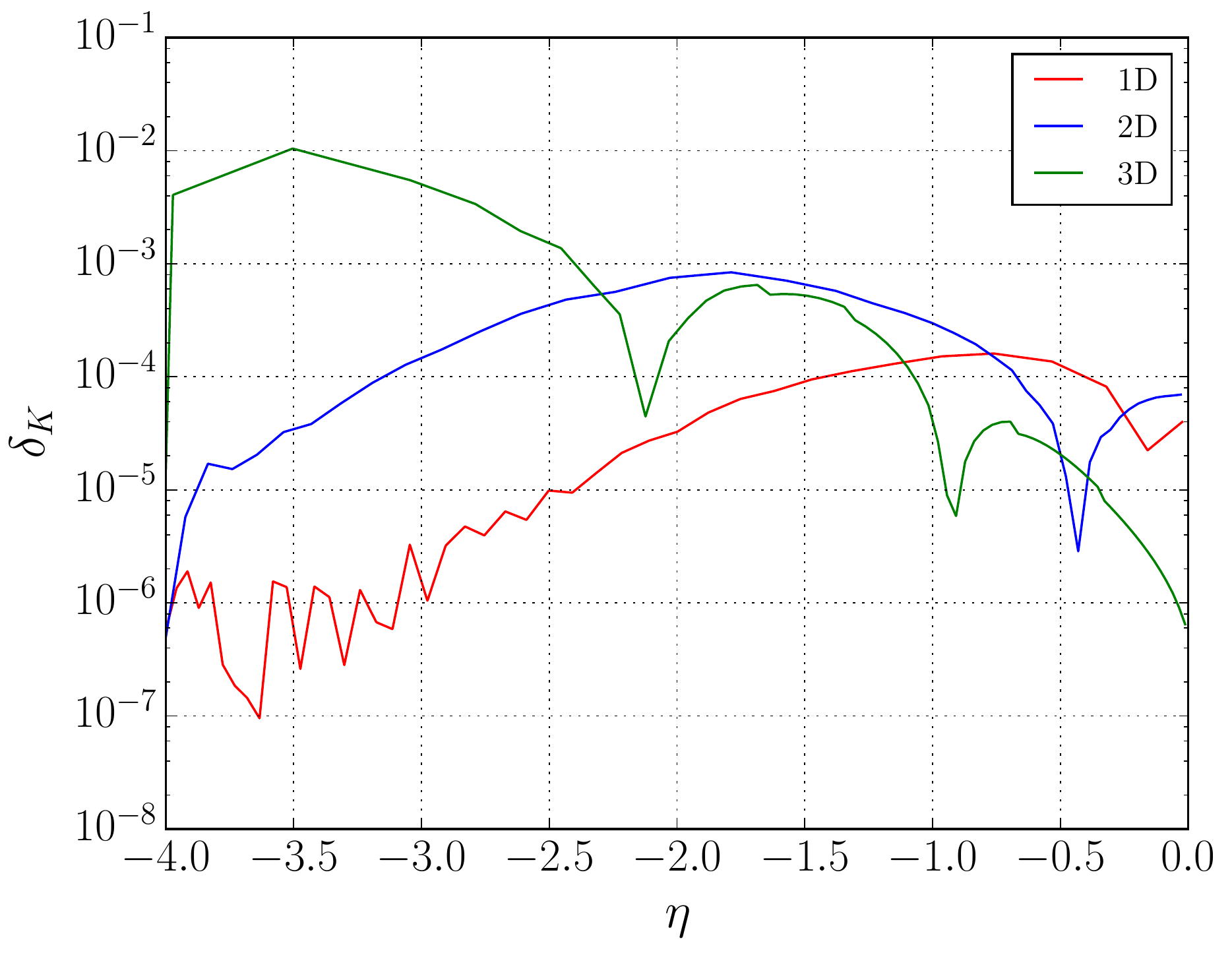}
\caption{\label{fig:deltaK}%
\small The plot shows $\delta_K$ as a function of time for simulations in 1D, 2D and 3D. The Layzer-Irvine test is passed ($\delta_K \ll 1$) in all cases.  See text for the parameters of the simulations.}
\end{figure}

As proposed in~\cite{Kopp:2017hbb} we also perform the Layzer-Irvine test of energy conservation in our simulations. In our setup, the kinetic and potential energies are naturally defined as 
\bea
K &=& -\frac{\kappa}2 \int \psi^* \Delta \psi \, , \\
W &=& \frac1{2} \int \bar V \,  \psi^* \psi \, .
\eea
Energy conservation is then spoiled by the explicit time dependence of $\kappa$ in $K$ and also in $\bar V$ (see Eq.~(\ref{eq:vbar})).
This yields the relation
\be
\partial_\eta (K+W) = -\frac12 K + \frac12 W \, .
\ee
This motivates the definition 
\be
\delta_K = \frac{\partial_\eta (K+W) - W/2 + K/2}{K} \, .
\ee
In Fig.~\ref{fig:deltaK} we show $\delta_K$ for 1D ($L=1000$~Mpc$/h$, $N = 16384$  and $\kappa_0 = 1$ Mpc$^2/h^2$), 2D ($L=1000$~Mpc$/h$, $N = 4096$  and $\kappa_0 = 1$ Mpc$^2/h^2$) and 3D ($L=1000$~Mpc$/h$, $N = 256$  and $\kappa_0 = 1$ Mpc$^2/h^2$) simulations. 

\section{Initialization redshift}\label{app_init}

\begin{figure}[ht]
\centering
  \includegraphics[width=0.45\textwidth]{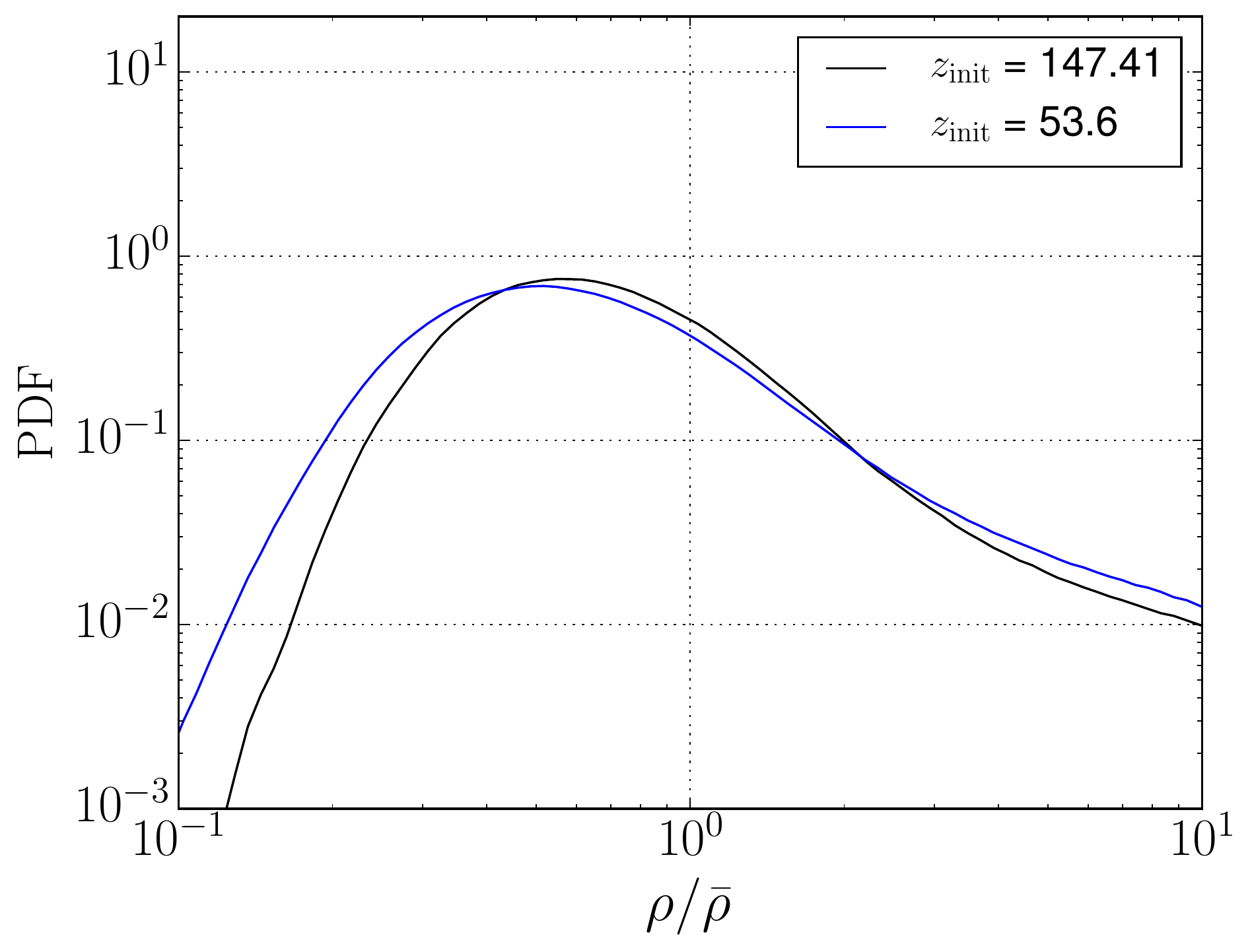} 
  \includegraphics[width=0.45\textwidth]{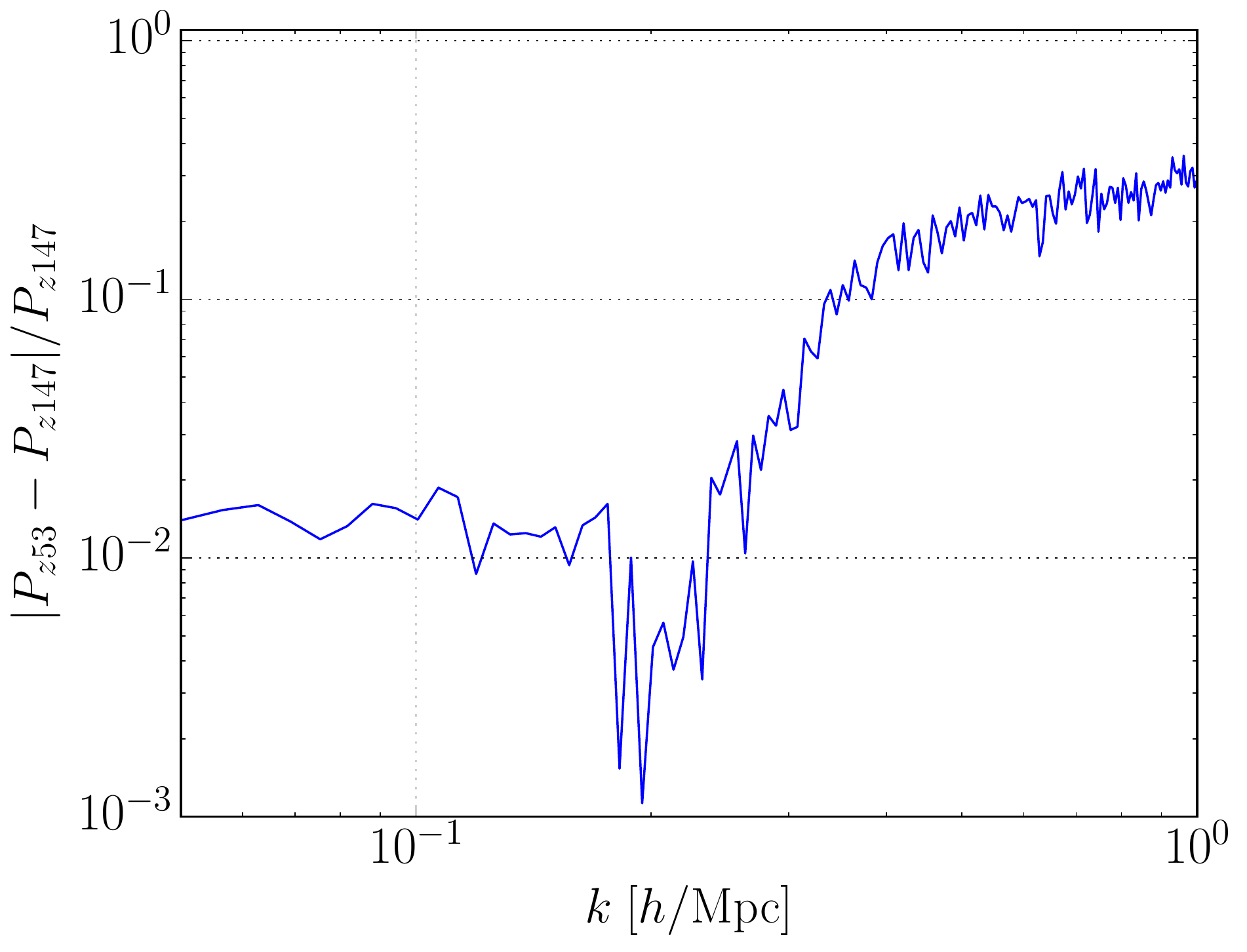}
\caption{\label{fig:init_z}%
\small Left: PDF for the matter density field at $z=0$ using redshifts $z_{\rm init} = 147.4$ and $z_{\rm init} = 53.6$.  Right: Relative difference of the matter power spectrum at $z=0$
obtained for the two  initial redshifts $z_{\rm init}$.}
\end{figure}

In figure~\ref{fig:init_z} we show the impact of initializing the SP evolution at two different redshifts $z_{\rm init} = 147.4$ and $z_{\rm init} = 53.6$ using $L=1000$~Mpc$/h$, $N = 8192$  and $\kappa_0 = 1$ Mpc$^2/h^2$. 
The initial redshift has a relatively strong influence on the PDF at $z=0$. 
The relative difference of the matter power spectrum is below $2\%$ for $k\lesssim 0.25h/$Mpc. $N$-body simulation results using Zel'dovich approximation as initial conditions also find similar discrepancies \cite{Schneider:2015yka}. 

\section{Computational time} \label{app:time}

In this appendix, we comment on the computational CPU time required for the \SP code.  
All the simulations were performed on the DESY Theory Cluster. 
For the Fourier transformations, we used the FFTW3 package \cite{FFTW05}. 

\begin{figure}[ht]
\centering
  \includegraphics[width=0.45\textwidth]{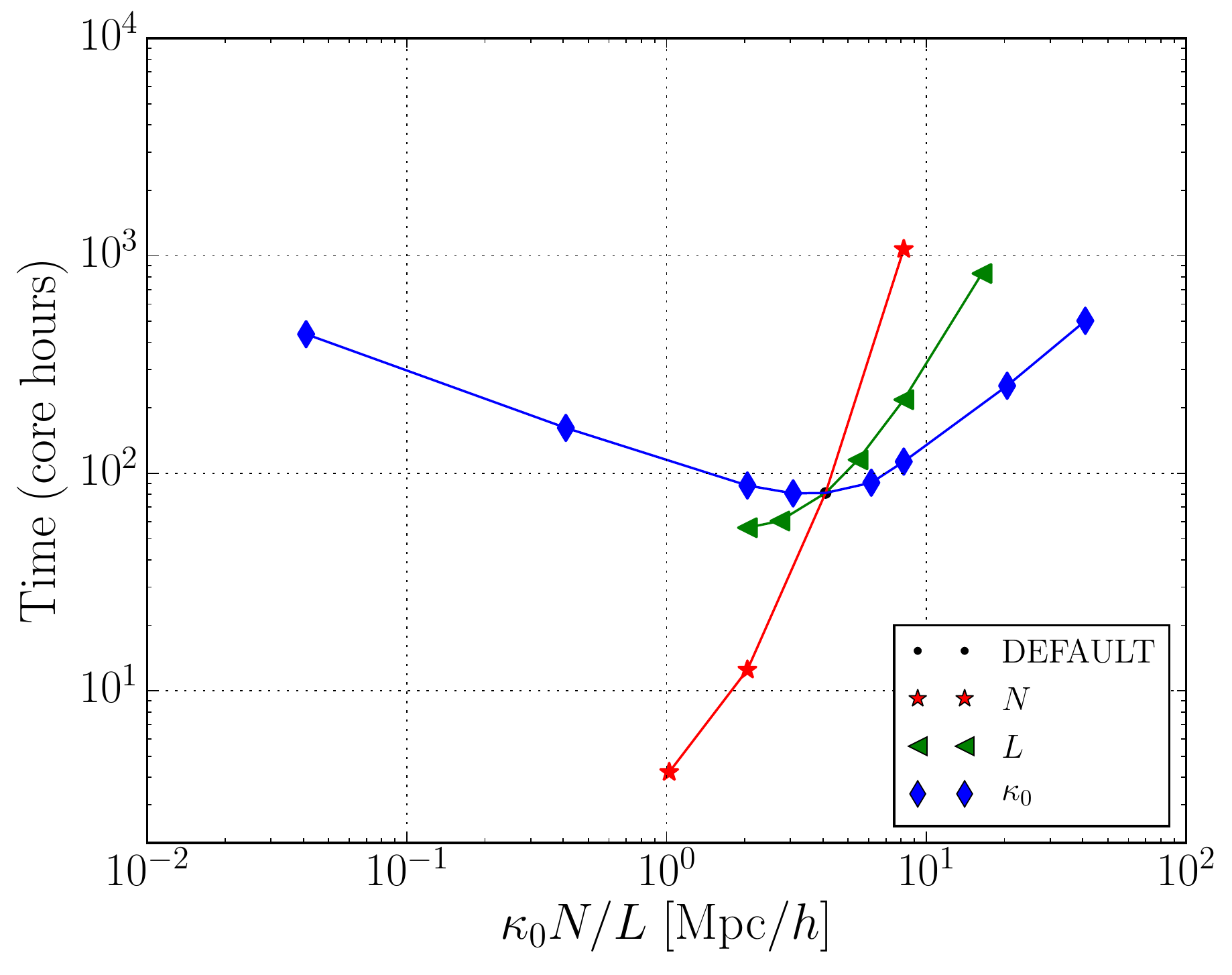}
\caption{\label{fig:Time_elaps}%
\small Dependence of the CPU core time required for the 2D \SP simulations on the simulation parameters $\kappa_0\propto\hbar$, $N$ and $L$.
The figure shows variations around the fiducial values.}
\end{figure}

In figure~\ref{fig:Time_elaps} we present the dependence of the simulation time (in core hours) on $\kappa_0$, $L$ and $N$ for the 2D case. Increasing $N$ has a twofold impact on the computational time: First, the time for each discrete Fourier transformation increases. Second, more non-linear scales are populated, increasing the argument of the potential rotations $U_V$.
This requires to decrease the time step $\Delta s$, as discussed in section~\ref{sec:sp_algo}. Reducing the box size $L$ also has similar effects on 
 $U_V$ (note that we use the combination $\kappa_0 N/L$ on the horizontal axis in figure~\ref{fig:Time_elaps}). 
Since $\arg(U_V) \propto 1/\kappa_0$ and $\arg(U_K) \propto \kappa_0$, extreme values of $\kappa_0$ also reduce the time step and correspondingly lead to an increase in computational time.

\section{The 1D case}\label{app:1d}

In this appendix, we present results for the one-dimensional case. As pointed out in the main text, even though the maximal possible
resolution in the 1D case is the highest, the (sampling) noise is very large due to the small number or modes. Nevertheless, we find it instructive
to consider the 1D case for studying the convergence when increasing the resolution. 

\begin{figure}[ht]
\centering
  \includegraphics[width=0.3\textwidth]{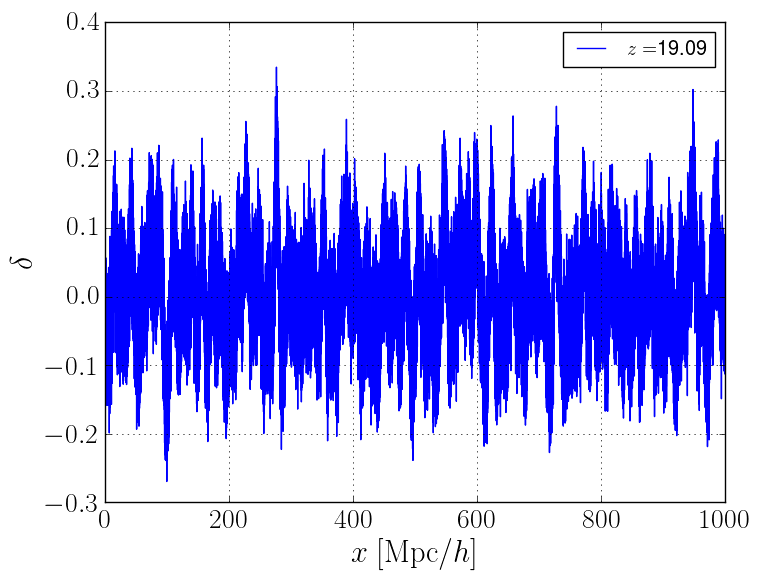}  
    \includegraphics[width=0.3\textwidth]{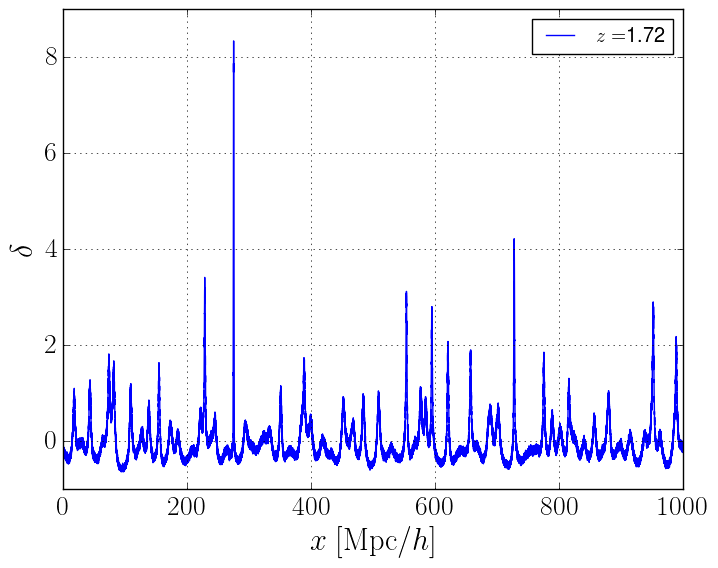}  
      \includegraphics[width=0.3\textwidth]{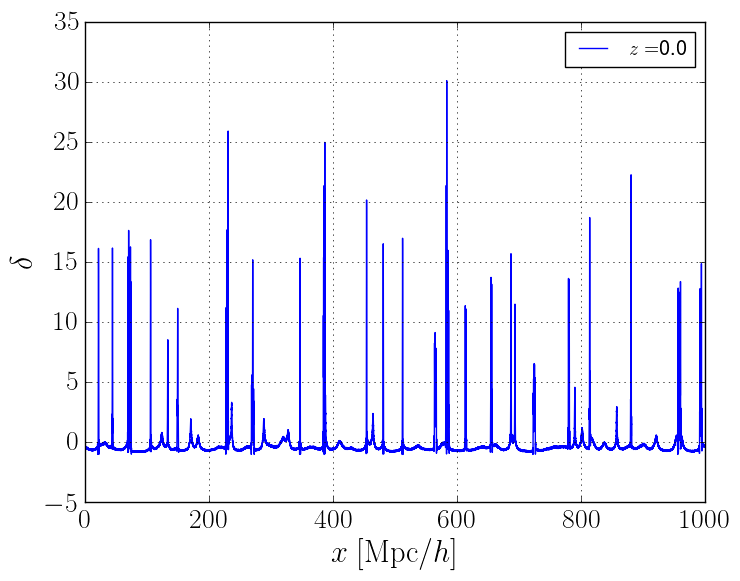}  
\caption{\label{fig:dens_1d_z}%
\small Overdensity field in the 1D SP system at three different redshifts. }
\end{figure}

In figure~\ref{fig:dens_1d_z} we show the overdensity field at three different redshifts, for a simulation with 
$L=1000$~Mpc/h, $\kappa_0 = 1\,{\rm Mpc}^2$ and $N = 2^{17}$ 
-- a substantial increase compared with both 2D and 3D cases. 
It is possible to see that a small initial fluctuation, for instance, at $x=250$~Mpc/h, evolves to form an overdense region. 

\begin{figure}[ht]
\centering
  \includegraphics[width=0.45\textwidth]{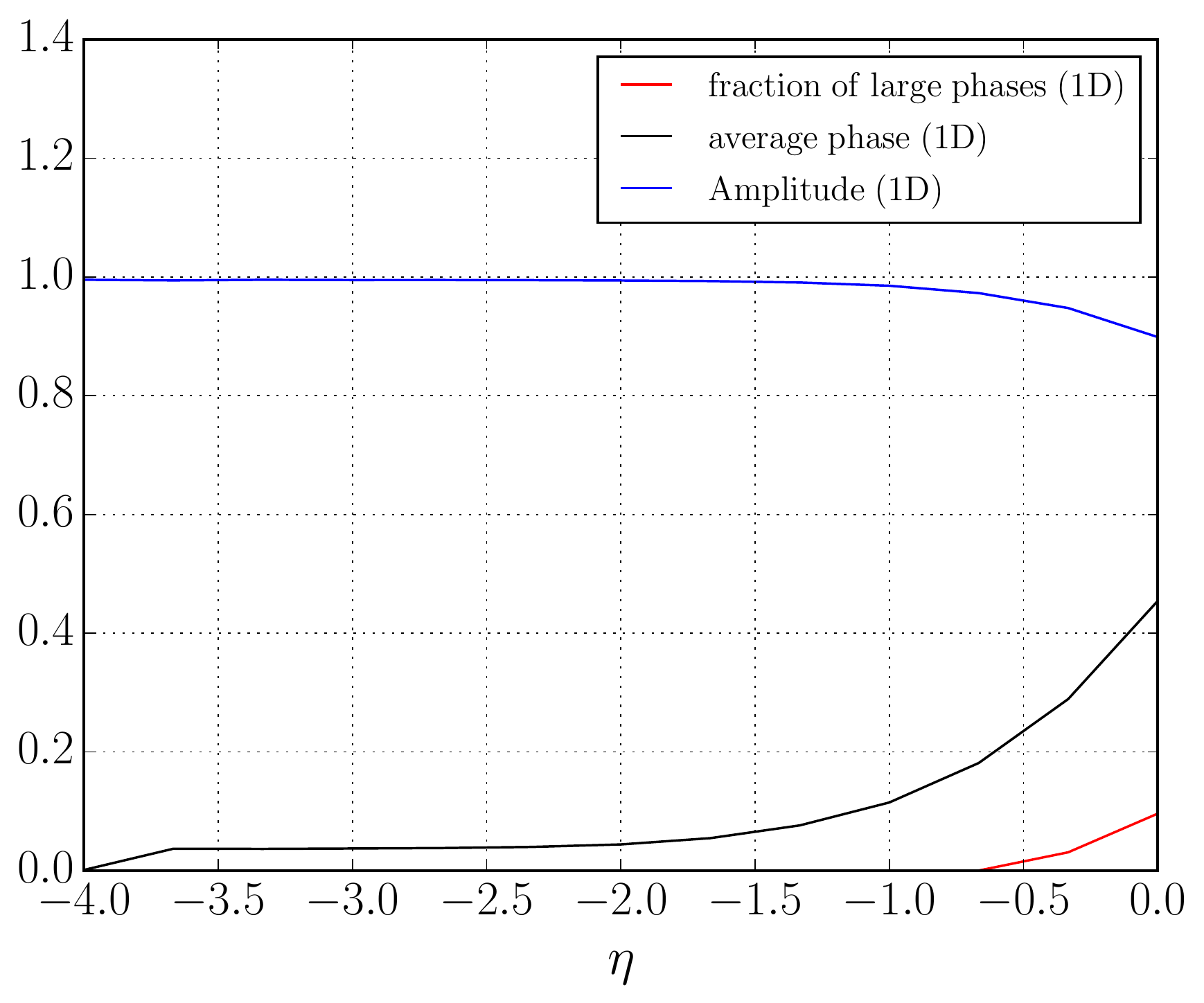}
  \includegraphics[width=0.45\textwidth]{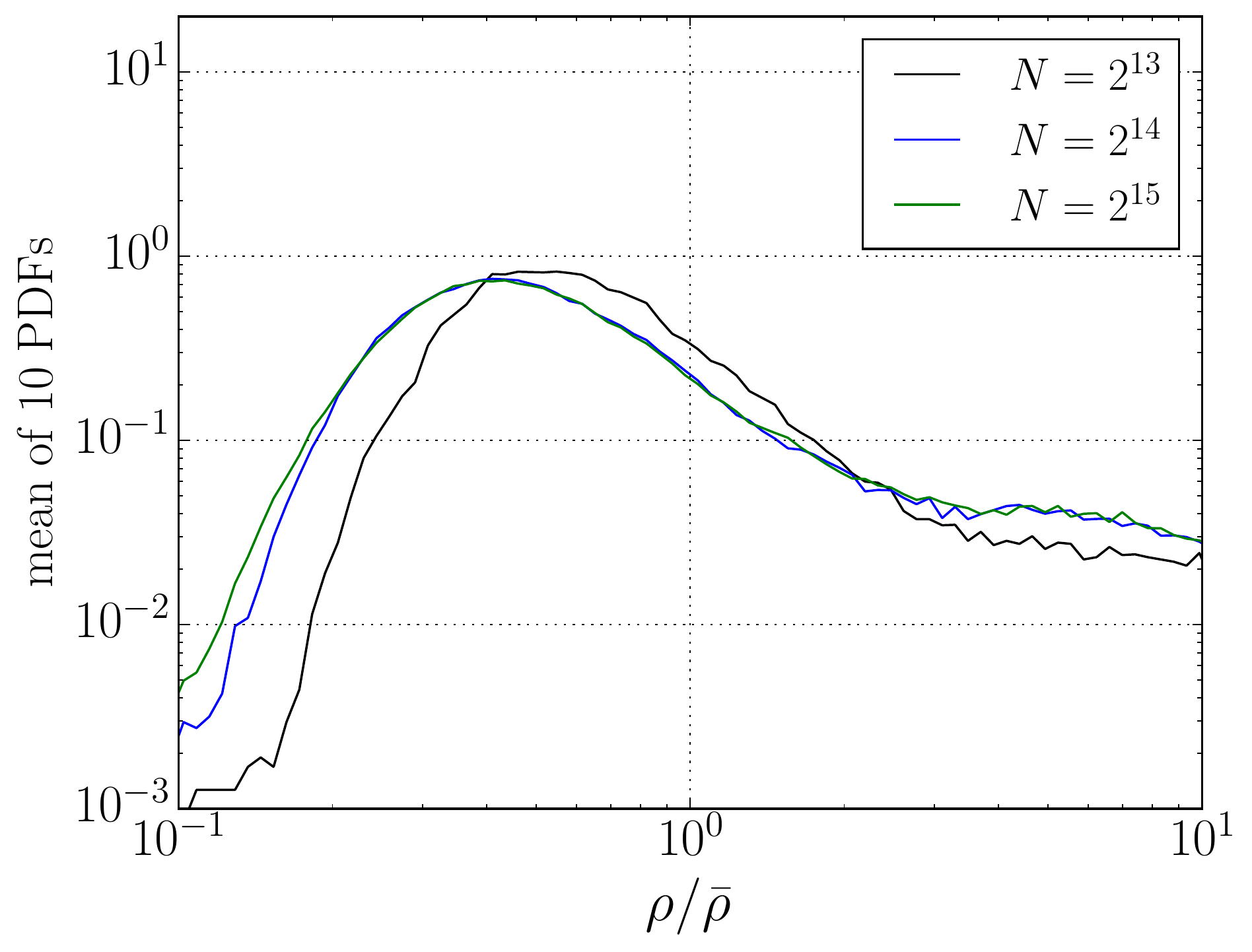}
\caption{\label{fig:pdfs_param_1d}%
\small Left: The plot shows the suppression factor $A^2$, the average of the relative phases and the fraction of large relative phases ($>\pi/4$) versus time (analogous to Fig.~\ref{fig:2dphases} for 2D). Right: PDF for different grid sizes $N$ averaged over 10 different initial conditions.}
\end{figure}

In the left panel of figure~\ref{fig:pdfs_param_1d}, we show the overall amplitude loss in the 1D case, defined as in equation~(\ref{eq:amp}) together with the averaged relative phases that indicate the failure of the grid to resolve the highest velocities. 
In the right panel of figure~\ref{fig:pdfs_param_1d}, we show the PDF obtained for different values of $N$. 
For $N \gtrsim 2^{14}$, the PDF starts to converge. 
We also run simulations with larger volume and larger $\kappa$ to confirm the reduction of a power loss in these cases.
We average over 10 different initial conditions to reduce noise and finite volume effects.

\bibliographystyle{utphys}
\bibliography{RefsSP}

\end{document}